\documentclass[11pt,longbibliography,preprintnumbers]{article}
\usepackage{cite}
\usepackage{scalerel}
\usepackage{amsmath,amsfonts,amssymb}
\usepackage{latexsym,epsfig}
\usepackage{dsfont}
\usepackage{extarrows}
\usepackage{hyperref}

\usepackage{comment} 

\def\hybrid{
        \topmargin -20pt
        \oddsidemargin 0pt
        \headheight 0pt \headsep 0pt
        \textwidth 6.25in 
        \textheight 9.5in 
        \marginparwidth .875in
        \parskip 5pt plus 1pt \jot = 1.5ex}

\hybrid

 \csname
@addtoreset\endcsname{equation}{section}

\def\Q{\mathds{Q}}
\def\cB{{\cal B}}

\def\cF{{\cal F}}
\def\cO{{\cal O}}
\def\cA{{\cal A}}
\def\cB{{\cal B}}
\def\cE{{\cal E}}
\def\cM{{\cal M}}
\def\cN{{\cal N}}
\def\cR{{\cal R}}

\def\cP{{\cal P}}

\def\cV{{\cal V}}

\def\cK{{\cal K}}

\def\cX{{\cal X}}
\def\cH{{\cal H}}

\def\del{\partial}
\def\l{\langle}
\def\r{\rangle}

\def\Tr{{\rm Tr}}

\def\bdel{\bar{\partial}}
\def\B{\square}
\def\Bb{\overline\square}

\allowdisplaybreaks

\def\bpm{\begin{pmatrix}}
\def\epm{\end{pmatrix}}
\newcommand{\bra}[1]{\langle#1\rvert}
\newcommand{\ket}[1]{\lvert#1\rangle}

\begin{document}

\begin{titlepage}
\rightline{}
\rightline{March 2022}
\rightline{HU-EP-22/07-RTG}  
\begin{center}
\vskip 1.5cm
{\Large \bf{The Gauge Structure of Double Field Theory \\[1ex]
follows from Yang-Mills Theory}}
\vskip 1.7cm

{\large\bf {Roberto Bonezzi, Felipe D\'iaz-Jaramillo and Olaf Hohm}}
\vskip 1.6cm

{\it  Institute for Physics, Humboldt University Berlin,\\
 Zum Gro\ss en Windkanal 6, D-12489 Berlin, Germany}\\[1.5ex] 
 ohohm@physik.hu-berlin.de, 
roberto.bonezzi@physik.hu-berlin.de, 
felipe.diaz-jaramillo@hu-berlin.de
\vskip .1cm

\vskip .2cm

\end{center}

\bigskip\bigskip
\begin{center} 
\textbf{Abstract}

\end{center} 
\begin{quote}

We show that to cubic order double field theory is encoded in Yang-Mills theory. 
To this end we use algebraic structures from string field theory as follows:
The $L_{\infty}$-algebra of Yang-Mills theory is the tensor product ${\cal K}\otimes \mathfrak{g}$ of 
the Lie algebra $\mathfrak{g}$ of the gauge group and a `kinematic algebra' ${\cal K}$ that is a 
$C_{\infty}$-algebra. This structure induces a cubic truncation 
of an $L_{\infty}$-algebra on the subspace of level-matched states of the  tensor product ${\cal K}\otimes \bar{\cal K}$ 
of two copies of the kinematic algebra.  
This $L_{\infty}$-algebra encodes double field theory.
More precisely, this construction relies on a particular form of the Yang-Mills $L_{\infty}$-algebra 
following  from  string field theory or  from the quantization of a suitable worldline theory.

\end{quote} 
\vfill
\setcounter{footnote}{0}
\end{titlepage}

\tableofcontents

\section{Introduction}

In this paper we show that, at least to cubic order, double field theory can be derived from  Yang-Mills theory through  
an off-shell and gauge invariant `double copy' construction. Double copy is a powerful tool in constructing 
 gravity scattering amplitudes from Yang-Mills or more general gauge theory 
 amplitudes  \cite{Bern:2008qj,Bern:2010ue,Bern:2019prr}. 
 It is a central pillar 
  of the modern amplitude program that in turn deemphasizes Lagrangians, off-shell states and gauge redundancies. 
 As such it is often considered to be beside the point to seek a Lagrangian understanding of double copy. 
Nevertheless, there have  been numerous  attempts to establish double copy relations at the level of 
a Lagrangian, see e.g.~\cite{Bern:1999ji,Bern:2010yg,Ananth:2007zy,Anastasiou:2018rdx,Borsten:2020xbt,Ferrero:2020vww,Tolotti:2013caa,Borsten:2020zgj,Beneke:2021ilf}. 
Recently, together with Plefka, two of us established in \cite{Diaz-Jaramillo:2021wtl} 
a close double copy relation to cubic order between the Lagrangians of 
Yang-Mills theory  and 
double field theory (DFT) \cite{Siegel:1993th,Hull:2009mi,Hull:2009zb,Hohm:2010jy,Hohm:2010pp}, which 
is a formulation of the string target space theory for graviton, B-field and dilaton that is T-duality invariant thanks 
to doubled coordinates.  
(See \cite{Aldazabal:2013sca,Berman:2013eva,Hohm:2013bwa,Tseytlin:1990va,Kugo:1992md,Siegel:1993xq} for reviews and  earlier work and \cite{Hohm:2011dz,Lee:2018gxc,Cho:2021nim} for previous work 
on double copy and DFT.) 
The construction of \cite{Diaz-Jaramillo:2021wtl}  requires integrating out the DFT dilaton 
and picking Siegel gauge at cubic order.   
Here we generalize  these results by showing that the full gauge invariant  DFT to cubic order, including all dilaton couplings,  follows from Yang-Mills theory. 

The  approach employed in this paper is algebraic, based on strongly homotopy algebras such as  $L_{\infty}$-algebras 
\cite{Zwiebach:1992ie,Lada:1994mn,Lada:1992wc}, which in turn are closely related to the Batalin-Vilkovisky (BV) formalism.   
$L_{\infty}$-algebras are generalizations of Lie algebras, defined on an integer graded vector space 
$\mathcal{X}=\bigoplus_{i\in \mathbb{Z}} X_i$ that encodes the space of fields, the space of gauge parameters, etc., 
and a potentially infinite series of graded symmetric maps or brackets $b_n$, $n=1,2,3,\ldots$, with $n$ inputs 
obeying quadratic generalized Jacobi  identities. 
When equipped with a graded symmetric inner product
the action of the theory is encoded in the $L_{\infty}$ brackets via 
 \begin{equation}\label{LinfintyYM}
S=\tfrac12\,\big\l A\,, b_1\big( A\big)\big\r+\tfrac{1}{3!}\,\big\l A\,, b_2\big(A,A\big)\big\r
+\tfrac{1}{4!}\,\big\l A\,, b_3\big(A, A, A\big)\big\r +\cdots \;, 
\end{equation}
where $A$ stands for all fields and the ellipsis indicates higher order terms. 
For the case of Yang-Mills theory in the standard formulation (i.e.~with at most quartic vertices) there are no higher 
brackets than 3-brackets. Similarly, gauge transformations, gauge algebra, Noether identities, etc., are 
all encoded in the $L_{\infty}$ structure (see \cite{Hohm:2017pnh} for  the general dictionary between 
$L_{\infty}$-algebras and field theory). 

It was shown by Zeitlin that the $L_{\infty}$-algebra of Yang-Mills theory can be viewed as a tensor 
product of the Lie algebra $\mathfrak{g}$ of the gauge group and a so-called $C_{\infty}$-algebra ${\cal K}$ \cite{Zeitlin:2008cc}:
 \begin{equation}\label{YMkinematic}
  L_{\infty} (\text{Yang-Mills})  \ =  \ {\cal K}\otimes \mathfrak{g}\;. 
 \end{equation} 
A $C_{\infty}$-algebra is a homotopy version of a differential graded commutative and associative algebra. 
It must be emphasized that (\ref{YMkinematic}) is in no way a symbolic relation but rather a completely precise statement 
about the tensor product of certain algebras.  
As the algebra ${\cal K}$ is obtained from the Yang-Mills $L_{\infty}$-algebra by `stripping off' color factors 
we will refer to it as the kinematic algebra of Yang-Mills theory, but we should point out that it does not immediately 
encode the relations that in the amplitude community are referred to as kinematic algebra (e.g.~\cite{Monteiro:2011pc,Bjerrum-Bohr:2012kaa,Ben-Shahar:2021zww}). 
In any case, double copy suggests that the tensor product ${\cal K}\otimes \bar{\cal K}$ encodes  a 
gravity theory.

In the following we will show, to cubic order in fields, 
that this progam can indeed be realized, with the gravity theory being DFT. 
This requires working with a particular formulation 
of Yang-Mills theory and  a subspace of ${\cal K}\otimes \bar{\cal K}$
corresponding to states satisfying the level-matching constraints of closed string theory. 
These features are directly motivated from the string theory origin of the double copy structure: 
the KLT relations between open string and closed string scattering amplitudes. 
As such, it is natural to suspect that string field theory (SFT) is the appropriate  framework  to 
make double copy manifest. Concretely, we will use an SFT inspired formulation of Yang-Mills theory 
with auxiliary fields so that all kinetic terms always come with a $\B$, and 
gauge fixing amounts to setting some auxiliary fields to zero. In BV language this formulation is known as 
non-minimal and can also be derived from the quantization of a worldline theory. 
We will then show that the $C_{\infty}$-algebra structure induces the (cubic truncation of the)  
$L_{\infty}$-algebra of DFT: 
\begin{equation}\label{DFTkinematic}
  L_{\infty} (\text{DFT})  \ =  \ \left[\,{\cal K} \, \otimes \, \bar{\cal K}\, \right]_{\text{level-matched}}\;, 
 \end{equation}
where the notation indicates the subspace of level-matched states.\footnote{A homotopy algebra approach to double copy was also developed in \cite{Borsten:2021hua}, but the outlined doubling procedure appears to be quite different.  See also \cite{Escudero:2022zdz, Reiterer:2019dys} for closely related applications.}

As a vector space, (\ref{DFTkinematic}) contains precisely the required objects to encode DFT. 
In particular, the doubling of coordinates of DFT is automatic and 
not imposed: given an algebra ${\cal K}$ of functions of $x$ and an independent algebra $\bar{\cal K}$ 
of functions of $\bar{x}$ 
the tensor product is a space of functions of $(x,\bar{x})$. This is a special case of 
the general relation that the tensor product of vector spaces ${\rm Fun}(M)$ of functions on a  manifold $M$ 
and functions on a second manifold $\bar{M}$ yields, 
under certain  topological assumptions, the algebra of functions on $M\times \bar{M}$: 
 \begin{equation}\label{tensoringfunctions}
  {\rm Fun}(M\times \bar{M}) = {\rm Fun}(M) \otimes {\rm Fun}(\bar{M})\;. 
 \end{equation}

It should be noted that establishing a homotopy algebra structure on a tensor product of such algebras 
in general is quite subtle. While algebras with no 
higher than 2-products or bracket, which are known as \textit{strict},  behave nicely under tensor products 
(so that, for instance, the tensor product of two strict $C_{\infty}$-algebras canonically yields a strict $C_{\infty}$-algebra), 
for general homotopy algebras it is more difficult to display a homotopy algebra structure on a tensor product. 
(See e.g.~\cite{Gaberdiel:1997ia}
for the case  of homotopy associative or $A_{\infty}$-algebras.)   
For a cubic truncation the $L_{\infty}$ structure requires only a 2-bracket with the correct graded symmetry 
properties obeying Leibniz relations with the differential. 
For both of these properties the truncation to the level-matched subspace in (\ref{DFTkinematic}) is instrumental. 
We do not yet know of a construction of the complete $L_{\infty}$-algebra on this space, 
as would be required in order to realize  double copy to all orders.

In order to appreciate the kind of detailed relations between Yang-Mills theory and DFT that are encoded 
in (\ref{DFTkinematic}) let us mention a particularly striking example: the DFT gauge transformations that are linear in 
the field $e_{\mu\bar{\mu}}$ (encoding metric and Kalb-Ramond fluctuations) are directly obtained 
from the 3-vertex of Yang-Mills theory! To see this note that according to (\ref{LinfintyYM}) 
the cubic term of Yang-Mills theory is  encoded in the 2-bracket 
on $\mathfrak{g}$-valued one-forms $A=A_{\mu}^{a}t_a dx^{\mu} $: 
 \begin{equation}\label{YM2Bracket}
  b_2(A,A)_{\mu}^{a} = f^{a}{}_{bc} (A^b \bullet  A^c)_{\mu}\;, 
 \end{equation}
where $f^{a}{}_{bc}$ are the structure constants of the Lie algebra $\mathfrak{g}$ (the `color factors') while $\bullet$
is a product on one-forms defined by 
 \begin{equation}\label{YMbullet}
  (v\bullet w)_{\mu} = v^{\nu}\partial_{\nu}w_{\mu} + (\partial_{\mu}v^{\nu}-\partial^{\nu}v_{\mu})w_{\nu}
  +(\partial_{\nu}v^{\nu})w_{\mu}-(v \leftrightarrow w)\;, 
 \end{equation}
where the Minkowski metric is used to raise and lower indices. (Of course, given this metric, we 
might as well view this as a  bracket of vectors 
rather than one-forms.) 
While the cubic term in the Yang-Mills action may be re-written in various equivalent ways, 
say by integrations by part,\footnote{For instance, in \cite{Fu:2016plh} it was observed that the cubic term can be written 
in terms of the conventional Lie bracket of vector fields.}
the corresponding $L_{\infty}$ 2-bracket (\ref{YM2Bracket}) is unambiguous: 
it takes, according to (\ref{YMbullet}), 
the form of an (antisymmetrized) generalized Lie derivative of DFT (with density weight one), 
just with the Minkowski metric instead of the $O(d,d)$ metric. However, in the perturbative formulation of DFT this 
is precisely the gauge transformation of $e_{\mu\bar{\mu}}$ with respect to the gauge parameter 
$\lambda_{\mu}$ \cite{Hull:2009mi}: 
 \begin{equation}\label{gaugeDFTe}
  \delta_{\lambda}^{(1)}e_{\mu\bar{\mu}} = \tfrac14\,(\lambda\bullet e_{\bar{\mu}})_{\mu} + \text{(auxiliary fields)}\;, 
 \end{equation} 
up to terms involving auxiliary fields of the SFT like formulation. In (\ref{gaugeDFTe}) the index 
$\bar{\mu}$ is viewed as inert, and there is an analogous gauge transformation for a gauge parameter 
$\bar{\lambda}_{\bar{\mu}}$, with all derivatives and indices in (\ref{YMbullet}) replaced by barred objects 
and  now with the index $\mu$ being inert. 
Note that the above relates a 2-bracket between fields on the Yang-Mills side to a 
2-bracket between field and gauge parameter on the DFT side, and  we will see that 
this precisely originates from the tensor product (\ref{DFTkinematic}) 
as  does the full cubic truncation of the $L_{\infty}$-algebra of DFT. 
Thus, this gauge invariant double copy is properly viewed as a map between the complete gauge theories 
as encoded in the corresponding $L_\infty$-algebras, 
as opposed to a simple redefinition between classical fields or their action.

The rest of this paper is organized as follows. In sec.~2 we write down the  $L_\infty$-algebra encoding Yang-Mills theory 
with a particular auxiliary field and identify the $C_{\infty}$ `kinematic algebra' ${\cal K}$. While this formulation was  inspired by open string field theory and was in fact  originally constructed by a related worldline quantization, 
we present our results without explicit reference to the SFT or  worldline formulation 
in order to keep this section self-contained. 
In sec.~3 we consider the subspace of ${\cal K}\otimes\bar{\cal K}$ of level-matched states and show that it inherits 
 the cubic truncation of an $L_\infty$-algebra that encodes DFT.  
 We close with a brief conclusion and outlook in sec.~4. 
For completeness, and since it is of interest in its own right,  we discuss  the worldline approach in an appendix.

\section{The Kinematic Algebra of Yang-Mills}
\label{sec:LooYM}

In this section we  present the $L_\infty$-algebra for Yang-Mills theory encoding  its gauge structure, classical field equations and Noether identities. We use a non-standard formulation that can be motivated by open string field theory or a worldline quantization. We then 
explain how `stripping off' the color part from the brackets one obtains a $C_\infty$-algebra, i.e.~a graded commutative algebra obeying associativity up to homotopy. This `kinematic algebra' will be used in the next section to derive double field theory to cubic order.

\subsection{$L_\infty$-algebra of Yang-Mills}

Let us consider the following form of the Yang-Mills action that includes an auxiliary scalar $\varphi$ in the free Lagrangian:
\begin{equation}\label{Classical cubic YM with phi}
S=\int dx\,\Tr\,\Big\{\tfrac12\,A^\mu \square A_\mu-\tfrac12\,\varphi^2+\varphi\,\partial_{\mu} A^{\mu} 
-\del_\mu A_\nu\,[A^{\mu},A^{\nu }]-\tfrac14\,[A^\mu, A^\nu]\,[A_\mu, A_\nu]
\Big\}\;,    
\end{equation}
where all fields are Lie algebra valued, \emph{e.g.}~$A_\mu=A_\mu^a\,t_a$, with Lie bracket $[\cdot, \cdot]$ and generators normalized as $\Tr\big(t_a t_b\big)=\delta_{ab}$. 
Upon integrating out $\varphi$ one recovers  the standard Yang-Mills action. This ensures that the standard cubic and quartic vertices in \eqref{Classical cubic YM with phi} are consistent with gauge invariance, provided that $A_\mu$ has the usual transformation rule and that $\varphi$ transforms as its on-shell value $\varphi=\partial_{\mu}A^{\mu}$, i.e.
\begin{equation}\label{easy gauge}
\delta A_\mu=\del_\mu\lambda+[A_\mu, \lambda]\;,\qquad\delta\varphi=\square \lambda+\del_\mu[A^\mu, \lambda]\;.    
\end{equation}
As we have mentioned, this non-standard form of the free action arises naturally from the BRST quantization of the $\cN=2$ spinning particle (see appendix \ref{worldline}), or from string field theory. 

We now describe  the $L_\infty$ structure of Yang-Mills in this formulation. An 
$L_\infty$ algebra is a graded vector space $\mathcal{X}=\bigoplus_{i} X_i$ 
endowed with multilinear maps $b_n$ of intrinsic degree $-1$, which obey quadratic Jacobi-like identities. 
Since Yang-Mills theory has at most quartic vertices the non-vanishing brackets are $b_1$, $b_2$ and $b_3$. The $L_\infty$ relations in this case are:
\begin{itemize}
\item Nilpotency of the differential
\begin{equation}
b_1^2 =0\;. 
\end{equation}
\item $b_1$ acts as a derivation on $b_2$ (Leibniz rule) 
\begin{equation}\label{Formal Leibniz}
b_1\big(b_2(x,y)\big)+b_2\big(b_1(x), y\big)+(-1)^xb_2\big(x,b_1(y)\big)=0\;. \end{equation}
\item Homotopy Jacobi identity
\begin{equation}\label{Formal Jacobi}
\begin{split}
&b_2\big(b_2(x,y),z\big)+(-1)^{yz}b_2\big(b_2(x,z),y\big)+(-1)^xb_2\big(x, b_2(y,z)\big)\\
&+b_1\big(b_3(x,y,z)\big)+b_3\big(b_1(x), y, z\big)+(-1)^{x}b_3\big(x, b_1(y), z\big)+(-1)^{x+y}b_3\big(x,y,b_1(z)\big)=0  \,. 
\end{split}    
\end{equation}
Here $x,y,z$ in exponents denote the $L_\infty$ degree of the corresponding element. 
\end{itemize}
Notice that, even if higher brackets vanish, one has higher relations involving $b_2b_3$ and $b_3b_2$ which we do not display here.
The sign conventions, referred to as the $b-$picture, are somewhat unconventional but are more convenient 
for field theory applications and also standard in string field theory \cite{Zwiebach:1992ie}. 
In the $b$-picture the brackets are graded symmetric: 
\begin{equation}
b_2(x,y)=(-1)^{xy}b_2(y,x)\;,\qquad b_3(x,y,z)=(-1)^{xy}b_3(y,x,z)\;, \quad {\rm etc.}
\end{equation}

Given this general structure, we now focus on Yang-Mills theory  defined by the action \eqref{Classical cubic YM with phi}.
The corresponding $L_\infty$ complex is the graded vector space
\begin{equation}
\mathcal{X}=\bigoplus_{i=-2}^{+1} X_i\;,    
\end{equation}
with differential $b_1$ of degree $-1$. The space of highest degree $X_1$ contains gauge parameters $\lambda$, while the spaces of lower degrees are identified as the space of fields, equations of motion and Bianchi/Noether identities, as shown in the following diagram:
\begin{equation}
\begin{array}{cccc}
X_1&\xlongrightarrow{b_1}\;X_0&\xlongrightarrow{b_1}\;X_{-1}&\xlongrightarrow{b_1}\;X_{-2}\\
\lambda&\hspace{7mm}\cA&\hspace{7mm}\cE&\hspace{7mm}\cN
\end{array}  \;.  
\end{equation}
The spaces of gauge parameters and Noether identities consist of scalars $\lambda$ and $\cN$, 
respectively, while
the space of fields $\cA$ and the space of field equations $\cE$ consist of doublets: 
$\cA=(A^\mu, \varphi)$ and $\cE=(E^\mu,  E)$.

\paragraph{Free theory and the differential $b_1$}

One can read off the differential $b_1$ from the linearized  field 
equations and gauge transformations of \eqref{Classical cubic YM with phi}:
\begin{equation}
b_1\big(\cA\big)=0\;,\qquad\delta\cA=b_1\big(\lambda\big)\;,   \end{equation}
to give  the action of the differential on $X_1$ and $X_0$: 
\begin{equation}\label{b1lambda b1A}
b_1\big(\lambda\big)=\bpm\del^\mu\lambda\\ \square\lambda\epm\in X_0\;,\qquad b_1\big(\cA\big)=\bpm\square 
A^\mu-\del^\mu\varphi\\\del\cdot A-\varphi\epm\in X_{-1}   \;, 
\end{equation}
where here and in the following we use the notation $\del\cdot A=\partial_{\mu}A^{\mu}$. 
Nilpotency of $b_1$ acting on $\lambda$ encodes gauge invariance of the free field equations, 
$b_1^2\big(\lambda\big)=b_1\big(\delta\cA\big)=0$. 
We define the differential $b_1$ acting on $\cE$ as
\begin{equation}
b_1\big(\cE\big)=\square E -\del_\mu E^\mu\;,
\end{equation}
in order to ensure that the free Noether identity is encoded in 
$b_1^2\big(\cA\big)=b_1\big(\cE\big)=0$. 
There is no further nontrivial realization of $b_1^2=0$. As we will discuss below, 
the differential $b_1$ coincides with the first-quantized BRST operator $Q$ of the associated  worldline theory.

\paragraph{Interacting theory and the brackets $b_2$ and $b_3$}

The field equations of the full theory can be written in $L_\infty$ form as
\begin{equation}\label{cubic eom}
b_1\big(\cA\big)+\tfrac12\,b_2\big(\cA, \cA\big)+\tfrac16\,b_3\big(\cA, \cA, \cA\big)=0 \;. 
\end{equation}
Since the auxiliary $\varphi$ does not enter interactions the 
two and three-brackets between fields take the form 
\begin{equation}
b_2\big(\cA,\cA\big)=\bpm b_2^\mu\big(A,A\big)\\0\epm\;,\qquad b_3\big(\cA,\cA,\cA\big)=\bpm b_3^\mu\big(A,A,A\big)\\0\epm\;,    
\end{equation}
for which one finds from the general dictionary between  $L_\infty$-algebras and field theory \cite{Hohm:2017pnh}:
\begin{equation}\label{b2b3 YM}
\begin{split}
b^\mu_2\big(A_1,A_2\big)&=2\,
\del_\nu\big[A^\nu_{(1}, A^\mu_{2)}\big]+2\,\big[f^{\mu\nu}_{(1},A_{2)\nu}\big]\;,\\
b^\mu_3\big(A_1, A_2, A_3\big)&=6\,\big[A_{\nu(1}, \big[A_2^\nu, A_{3)}^\mu\big]\big]\;, 
\end{split} 
\end{equation}
with the abelian field strength $f_{\mu\nu}=\del_\mu A_\nu-\del_\nu A_\mu$. 
The field equations \eqref{cubic eom} are covariant under the deformed gauge transformation
\begin{equation}
\delta\cA=b_1\big(\lambda\big)+b_2\big(\lambda, \cA\big)  \;, 
\end{equation}
which, comparing with \eqref{easy gauge}, fixes the two bracket between a field and a gauge parameter to be
\begin{equation}\label{classical b2Alambda}
b_2\big(\lambda, \cA\big)=\bpm
[A^\mu, \lambda]\\\del_\nu[A^\nu, \lambda]
\epm  \;.  
\end{equation}
Demanding that the Leibniz property \eqref{Formal Leibniz} holds for all the allowed combinations of inputs, one finds the following list of non-vanishing brackets:
\begin{equation}\label{allb2s}
\begin{split}
b_2\big(\lambda_1, \lambda_2\big)&=-[\lambda_1, \lambda_2]
\quad
\hspace{44mm}b_2\big(\cA, \lambda\big)=\bpm[A^\mu, \lambda]\\\del_\nu[A^\nu, \lambda]\epm
\;,\\
b_2\big(\cA_1, \cA_2\big)&=2\bpm\del_\nu\big[A_{(1}^\nu, A_{2)}^\mu\big]+\big[f_{(1}^{\mu\nu},A_{2)\nu}\big]\\0\epm\;,\qquad
b_2\big(\lambda, \cE\big)=-\bpm[\lambda, E^\mu-\del^\mu E]\\0\epm
\;,\\
b_2\big(\cA, \cE\big)&=-[A_\mu, E^\mu-\del^\mu E]\;,\quad\hspace{26mm} b_2\big(\lambda, \cN\big)=-[\lambda, \cN]
\;.
\end{split}    
\end{equation}
The homotopy Jacobi identities \eqref{Formal Jacobi} are satisfied with the only non-vanishing three-bracket $b^\mu_3(A_1,A_2,A_3)$ given in \eqref{b2b3 YM}.

\paragraph{Inner product and action}

We conclude the discussion of the $L_\infty$ algebra of Yang-Mills by giving the inner product, which allows us to write the action in the standard $L_\infty$ form \cite{Zwiebach:1992ie, Hohm:2017pnh}.
Specifically, 
the inner product of $\cX$ is a degree $+1$ pairing, including a map between fields and field equations:
\begin{equation}
\l\;,\;\r\;:X_0\times X_{-1}\;\longrightarrow\;\mathbb{R}\;.
\end{equation}
Given a field $\cA$ and a field equation $\cE$ with components
\begin{equation}
\cA=\bpm A^\mu\\\varphi\epm\in X_0\;,\quad \cE=\bpm E^\mu\\E\epm\in X_{-1}\;, \end{equation}
we define their inner product by the spacetime integral
\begin{equation}\label{inner YM}
\l\cA, \cE\r=\int dx\,\Tr\Big\{A^\mu E_\mu+\varphi\,E\Big\}\;,  \end{equation}
with a similar pairing between gauge parameters $\lambda$ and Noether identities ${\cal N}$. 
Using the form \eqref{b1lambda b1A} and \eqref{b2b3 YM} for the brackets one may verify  that the Yang-Mills action \eqref{Classical cubic YM with phi} can be written in terms of the inner product as 
\begin{equation}
S=\tfrac12\,\big\l\cA\,, b_1\big(\cA\big)\big\r+\tfrac{1}{3!}\,\big\l\cA\,, b_2\big(\cA,\cA\big)\big\r+\tfrac{1}{4!}\,\big\l\cA\,, b_3\big(\cA, \cA, \cA\big)\big\r \;. 
\end{equation}
The Euler-Lagrange equations of this action indeed take  the form \eqref{cubic eom}.

\subsection{The Kinematic Algebra}

Having presented the $L_\infty$ algebra of Yang-Mills, our next goal will be to disentangle the color degrees of freedom from the elements of $\cX$ and from the brackets. This defines the vector space $\cK$ of the kinematic algebra, whose elements are spacetime fields with no color dependence. Similarly, we will define purely kinematic products $m_n$ acting on the kinematic vector space. 

Let us start with the vector space itself. Since all elements of the $L_\infty$ algebra $\cX$ are Lie algebra-valued fields, they can be written as
$x=x^a\,t_a\in \cX$ in terms of generators $t_a\in \mathfrak{g}$, where $\mathfrak{g}$ is the Lie algebra of the gauge group. This shows that the $L_\infty$ complex $\cX$ has the structure of a tensor product space $\cX=\cK\otimes\mathfrak{g}$, where $\mathfrak{g}$ is endowed with the standard Lie bracket, and $\cK$ is the vector space of the kinematic algebra, such that $t_a\in \mathfrak{g}$ and $x^a\in \cK$. 
If one considers the Lie algebra $\mathfrak{g}$ as a special case of an $L_\infty$-algebra concentrated in degree $+1$ (in the $b-$picture), the generators $t_a$ have degree $+1$, while the Lie bracket $[\;,\;]$  has intrinsic degree $-1$. Next, we shall define the degrees of the kinematic algebra by declaring that the degrees of elements and maps of $\cK$ and $\mathfrak{g}$ are additive, so that
\begin{equation}
x=u\otimes t\;,\quad u\in\cK\;,\;t\in\mathfrak{g}\quad\rightarrow\quad |x|_\cX=|u|_\cK+|t|_\mathfrak{g}=|u|_\cK+1\;.
\end{equation}
This implies that the kinematic algebra $\cK$ is given by the direct sum
\begin{equation}
\cK=\bigoplus_{i=-3}^0K_i\;. 
\end{equation}
Although elements of $\cK$ are spacetime fields with no color degrees of freedom left, informally we still refer to 
$K_0$ as the space of gauge parameters $\lambda$, $K_{-1}$ as the space of fields $\cA=(A^\mu, \varphi)$, etc., 
i.e.~we keep the same symbols for elements  of $\cK$ in order to  avoid burdening the notation. 
We shall now define the multilinear maps $m_n$ on $\cK$. 

\paragraph{Differential $m_1$} 
Since the Lie algebra $\mathfrak{g}$ plays no role in defining the free field equations, the differential $m_1$ is the same as the $L_\infty$ differential $b_1$ or, more precisely, 
\begin{equation}\label{m1 formal}
b_1(x)=b_1(u\otimes t)=m_1(u)\otimes t\;,\quad |m_1|=-1\;,  
\end{equation}
where the degree of $m_1$ can be inferred from the definition \eqref{m1 formal} and $|b_1|=-1$.
This yields the explicit realization
\begin{equation}\label{allm1s}
m_1\big(\lambda\big)=\bpm\del^\mu\lambda\\\square\lambda\epm\in K_{-1}\,,\quad m_1\big(\cA\big)=\bpm\square A^\mu-\del^\mu\varphi\\\del\cdot A-\varphi\epm\in K_{-2}\,,\quad m_1\big(\cE\big)=\square E-\del_\mu E^\mu\in K_{-3}\,,  
\end{equation}
and the proof of  $m_1^2=0$ is identical to the proof  of $b_1^2=0$.

\paragraph{Two-product $m_2$}
We shall now define a degree zero graded commutative product $m_2$ on $\cK$, which thus obeys 
\begin{equation}\label{m2 sym}
m_2(u_1, u_2)=(-1)^{u_1u_2}m_2(u_2, u_1)\;, \qquad |m_2|=0\;. 
\end{equation}
Given two vectors $x_1, x_2\in \cX$ of the form $x_i=u_i\otimes t_i$ with $u_i\in\cK$ and $t_i\in\mathfrak{g}$, the $L_\infty$ bracket $b_2$ can be written as
\begin{equation}\label{b2m2 formal}
b_2(x_1,x_2)=b_2(u_1\otimes t_1, u_2\otimes t_2)=(-1)^{x_1}m_2(u_1, u_2)\otimes[t_1, t_2]\;,
\end{equation}
which serves as an implicit  definition of $m_2$.
The degree $|m_2|=0$ is compatible with \eqref{b2m2 formal} since 
$|b_2|=-1$ and $|[\;,\;]|=-1$. The sign factor $(-1)^{x_1}$ has been chosen so that \eqref{b2m2 formal} is also compatible with the symmetry property of $b_2$.
This can be checked by computing
\begin{equation}
\begin{split}
b_2(x_2,x_1)&=(-1)^{x_2}m_2(u_2, u_1)\otimes[t_2, t_1]=(-1)^{u_2+1}m_2(u_2, u_1)\otimes[t_2, t_1]\\
&=(-1)^{u_1u_2+u_2}m_2(u_1, u_2)\otimes[t_1, t_2]=(-1)^{u_1u_2+u_2+u_1+1}b_2(x_1, x_2)\\
&=(-1)^{x_1x_2}b_2(x_1, x_2)\;.
\end{split}    
\end{equation}
Let us point out that the definition \eqref{b2m2 formal} does not imply the graded symmetry of $m_2$. Rather, it implies that any part of $m_2$ which is not graded symmetric is projected out from the resulting $b_2$. The symmetry property \eqref{m2 sym} has thus to be seen as part of the definition of $m_2$.

Let us now use the definition \eqref{b2m2 formal} to give the explicit form of the Yang-Mills two-products $m_2$.
It is convenient to use the decomposition along a basis $t_a$ of $\mathfrak{g}$, such that $x=x^a\otimes t_a$. In this case \eqref{b2m2 formal}
reduces to
\begin{equation}\label{b2m2 components}
b_2(x, y)=(-1)^xf^a{}_{bc}\,m_2(x^b, y^c)\otimes t_a\;,  
\end{equation}
in terms of component fields with adjoint color indices. Using \eqref{b2m2 components} and the brackets \eqref{allb2s} one finds that the non-vanishing products are given by
\begin{equation*}
\begin{split}
m_2\big(\lambda_1, \lambda_2\big)&=\lambda_1 \lambda_2\;\in \; K_0\;,\\[2mm]
m_2\big(\cA, \lambda\big)&=\bpm A^\mu \lambda\\[2mm]\del_\nu(A^\nu \lambda)\epm\;\in \; K_{-1}\;,    
\end{split}    
\end{equation*}

\begin{equation}\label{allm2s}
\begin{split}
m_2\big(\cA_1, \cA_2\big)&=\bpm\big(A_1\bullet A_2\big)^\mu\\0\epm
\;\in \; K_{-2}\;,\\[2mm]
m_2\big(\lambda, \cE\big)&=\bpm\lambda( E^\mu-\del^\mu E)\\0\epm\; \in\;  K_{-2}\;,\\[2mm]
m_2\big(\cA, \cE\big)&=-A_\mu( E^\mu-\del^\mu E)\; \in \; K_{-3}\;,\\[2mm] m_2\big(\lambda, \cN\big)&=\lambda\, \cN\;\in \; 
K_{-3}\;,
\end{split}    
\end{equation}
where the antisymmetric product $\bullet$ between vector fields, already defined in the introduction, is given by 
\begin{equation}\label{bullet}
\big(V\bullet W\big)^\mu=V^\nu\del_\nu W^\mu+\big(\del^\mu V_\nu-\del_\nu V^\mu\big)W^\nu+\big(\del_\nu V^\nu\big)W^\mu-\big(V\leftrightarrow W\big) \;.   
\end{equation}
The non-diagonal products with a different order of inputs, \emph{e.g.}~$m_2\big(\lambda, \cA\big)$ or $m_2\big(\cE, \cA\big)$, are defined by the symmetry property \eqref{m2 sym}.

\paragraph{Three-product $m_3$} 

The only non-vanishing three-bracket of Yang-Mills theory acts on three degree zero elements, which are  the vector components of three fields: $b^\mu_3\big(A_1,A_2,A_3\big)$. For this reason, the only non-vanishing kinematic three-product $m_3(u_1, u_2, u_3)$ acts on three elements $u_i\in K_{-1}$ of degree $-1$. 
We thus define a three-product $m_3 :\,K_{-1}^{\otimes 3}\,\rightarrow\,K_{-2}$ of degree $|m_3|=+1$, acting as
\begin{equation}\label{m3}
\begin{split}
m_3\big(\cA_1, \cA_2, \cA_3\big)&=\bpm m^\mu_3\big(A_1, A_2,A_3\big) \\0\epm \;,\\
m^\mu_3\big(A_1, A_2,A_3\big)&=A_1\cdot A_2\,A_3^\mu+A_3\cdot A_2\,A_1^\mu-2\,A_1\cdot A_3\,A_2^\mu\;.
\end{split}
\end{equation}
Upon inspection one may verify that this  obeys 
\begin{equation}\label{m3 symmetries}
m_3\big(\cA_1, \cA_2, \cA_3\big)=m_3\big(\cA_3, \cA_2, \cA_1\big)\;,\qquad m_3\big(\cA_{(1}, \cA_2, \cA_{3)}\big)=0 \;.  
\end{equation}
In the language of Young tableaux this means that $m_3$ has the 
symmetry property of a $(2,1)$ `hook' Young tableau. Consider now three degree zero elements $x_1, x_2,x_3\in X_0$ in the $L_\infty$ algebra $\cX$. Taking them to be of the form $x_i=u_i\otimes t_i$, with $t_i\in \mathfrak{g}$ and $u_i\in K_{-1}$, the three-bracket \eqref{b2b3 YM} can be written as
\begin{equation}\label{b3m3 formal}
\begin{split}
b_3(x_1,x_2,x_3)&=b_3(u_1\otimes t_1,u_2\otimes t_2,u_3\otimes t_3)\\
&=m_3(u_1, u_{2}, u_{3})\otimes[t_1,[t_2,t_3]]+m_3(u_2, u_{1}, u_{3})\otimes[t_2,[t_1,t_3]]\;,
\end{split}    
\end{equation}
which guarantees total symmetry in the inputs $x_1$, $x_2$ and $x_3$ thanks to \eqref{m3 symmetries} and the symmetry properties of the nested Lie algebra bracket.
In order to recognize the previous  expression \eqref{b2b3 YM}, it is convenient to use the standard basis decomposition $x=x^a\otimes t_a$, which yields
\begin{equation}
b^\mu_3\big(A_1,A_2,A_3\big)=2\,f^a{}_{be}f^e{}_{cd}\,m^\mu_3\big(A^b_{(1}, A^c_{2}, A^d_{3)}\big)\otimes t_a \;.   
\end{equation}
One can use \eqref{m3} in the expression above to immediately recover \eqref{b2b3 YM}.

\subsubsection*{$C_\infty$-algebra}\label{sec:Cinftyrelations}

In the last part of this section we will prove that the vector space $\cK$ endowed with the products 
$m_1$, $m_2$ and $m_3$ given by \eqref{allm1s}, \eqref{allm2s} and \eqref{m3} defines a $C_\infty$-algebra: 
a graded vector space endowed with multilinear products $m_n$ of degree $|m_n|=n-2$ obeying certain symmetry properties and associativity up to homotopy, see e.g. \cite{MARKL1992141,kadeishvili2002structure,AAA2008}. Specifically, $C_\infty$-algebras are 
$A_\infty$-algebras with products that obey additional symmetry properties. In particular, the two and three-products $m_2$ and $m_3$ obey
\begin{equation}
\begin{split}
m_2(u_1,u_2)&=(-1)^{u_1u_2}\,m_2(u_2,u_1) \;,\\
m_3(u_1, u_2, u_3)&=(-1)^{u_2u_3}\,m_3(u_1, u_3, u_2)-(-1)^{(u_1+u_2)u_3}\,m_3(u_3, u_1, u_2)\;,
\end{split}    
\end{equation}
which is the case for the Yang-Mills products \eqref{allm2s} and \eqref{m3}, recalling that $m_3$ is non-vanishing only for $|u_1|=|u_2|=|u_3|=-1$. The  quadratic relations for the $m_n$ are as follows: First, the differential $m_1$ is nilpotent:
\begin{equation}
m_1^2(u)=0\;,\quad |m_1|=-1\;,    
\end{equation}
and acts as a derivation with respect to $m_2$ (Leibniz rule): 
\begin{equation}\label{m2 formal Leibniz}
m_1\big(m_2(u_1, u_2)\big)=m_2\big(m_1(u_1), u_2\big)+(-1)^{u_1}\,m_2\big(u_1,m_1(u_2)\big)\;,\quad |m_2|=0\;. 
\end{equation} 
Second, one has associativity of $m_2$ up to homotopy, which is expressed as
\begin{equation}\label{m2m3 Aoo}
\begin{split}
&m_2\big(m_2(u_1,u_2),u_3\big)-m_2\big(u_1,m_2(u_2,u_3)\big)=m_1\big(m_3(u_1,u_2,u_3)\big)\\
&+m_3\big(m_1(u_1),u_2,u_3\big)+(-1)^{u_1}\,m_3\big(u_1,m_1(u_2),u_3\big)+(-1)^{u_1+u_2}\,m_3\big(u_1,u_2,m_1(u_3)\big) \;,     
\end{split}  
\end{equation}
where $m_3$ has degree $+1$. 
Even though there are no higher products, $m_2$ and $m_3$ have to satisfy \cite{Borsten:2021hua}
\begin{equation}\label{consistent m3}
\begin{split}
m_2\big(m_3(u_1, u_2, u_3), u_4\big)&+(-1)^{u_1}m_2\big(u_1, m_3(u_2, u_3, u_4)\big)=m_3\big(m_2(u_1,u_2), u_3, u_4\big)\\
&-m_3\big(u_1,m_2(u_2, u_3), u_4\big)+m_3\big(u_1,u_2, m_2(u_3, u_4)\big)\;,
\end{split}    
\end{equation}
in order to be consistent with the absence of $m_4$. The last possible relation, involving $m_3m_3$, vanishes identically for degree reasons.
We will now show that the differential \eqref{allm1s}, two and three-products \eqref{allm2s} and \eqref{m3} obey the graded Leibniz and homotopy associativity relations. Finally, we will establish \eqref{consistent m3}, which concludes the proof of the quadratic relations.

Nilpotency of the differential $m_1$ follows immediately as above. For the Leibniz property we shall proceed in order, starting from the highest degree: 

\paragraph{Degree $-1$} The only Leibniz relation at degree $-1$ involves the product of two gauge parameters $\lambda_1$ and $\lambda_2$, and reads
\begin{equation}
\begin{split}
m_1\big(m_2(\lambda_1,\lambda_2)\big)&=\bpm\del^\mu(\lambda_1\lambda_2)\\\square(\lambda_1\lambda_2)\epm    =\bpm\del^\mu\lambda_1\lambda_2+\lambda_1\del^\mu\lambda_2\\
\square\lambda_1\lambda_2+\lambda_1\square\lambda_2+2\,\del^\nu\lambda_1\del_\nu\lambda_2\epm\\
&=\bpm(\del^\mu\lambda_1)\lambda_2\\\del_\nu(\del^\nu\lambda_1\lambda_2)\epm+\bpm(\del^\mu\lambda_2)\lambda_1\\\del_\nu(\del^\nu\lambda_2\lambda_1)\epm\\
& =m_2\big(m_1(\lambda_1), \lambda_2\big)+m_2\big(\lambda_1, m_1(\lambda_2)\big)\;,
\end{split}   
\end{equation}
as can be seen from \eqref{allm1s} and \eqref{allm2s}. In the $L_\infty$-algebra associated to $\cK$ by tensoring with $\mathfrak{g}$, this relation encodes closure of the gauge transformations of Yang-Mills.

\paragraph{Degree $-2$} Also in degree $-2$ one has only one relation, between a gauge parameter $\lambda$ and a field $\cA$:
\begin{equation}
\begin{split}
m_1\big(m_2(\cA, \lambda)\big)&=\bpm\square(A^\mu\lambda)-\del^\mu\del_\nu(A^\nu\lambda) \\\del_\nu(A^\nu\lambda)-\del_\nu(A^\nu\lambda)\epm=\bpm(\square A^\mu-\del^\mu\del\cdot A)\lambda\\0\epm\\
&+\bpm A^\mu\square\lambda-A^\nu\del_\nu\del^\mu\lambda+2\,\del^\nu\! A^\mu\,\del_\nu\lambda-\del^\mu\! A^\nu\,\del_\nu\lambda-\del\cdot A\,\del^\mu\lambda\\0\epm\\
&=m_2\big(m_1(\cA), \lambda\big)-m_2\big(\cA, m_1(\lambda)\big)\;,
\end{split}    
\end{equation}
which corresponds to gauge invariance of the field equations in $\cX$.

\paragraph{Degree $-3$}

The relations in lowest degree can either take two fields $\cA_1$ and $\cA_2$, thus expressing the deformation of the Noether identity $\cN$, or a gauge parameter $\lambda$ and a field equation $\cE$, which corresponds to the consistency of $\cN=0$ with gauge symmetries.
The first one yields
\begin{equation}
\begin{split}
m_1\big(m_2(\cA_1, \cA_2)\big)&=2\,\del_\mu\Big(\del\cdot A_{[2}A^\mu{}_{1]}+2\,A^\nu_{[2}\del_\nu A^\mu{}_{1]}+\del^\mu A^\nu_{[2} A_{1]\nu}\Big)\\
&=2\,\Big(\del_\mu\del\cdot A_{[2}A^\mu{}_{1]}+2\,A^\nu_{[2}\del_\nu\del\cdot A{}_{1]}+\square A^\nu_{[2} A_{1]\nu}\Big)\\
&=-2\,A^\mu_{[2}\,\Big(\square A_{1]\mu}-\del_\mu\del\cdot A_{1]}\Big)=2\,m_2\big(m_1(\cA_{[1}), \cA_{2]}\big)\;,
\end{split}    
\end{equation}
where we point out that, given a field equation $\cE=m_1(\cA)=\big(\B A^\mu-\del^\mu\varphi\;,\; \del\cdot A-\varphi\big)$, the combination $E^\mu-\del^\mu E=\B A^\mu-\del^\mu\del\cdot A$ is the usual Maxwell equation not involving $\varphi$.
The last Leibniz relation, with $\lambda$ and $\cE$ as inputs, gives
\begin{equation}
\begin{split}
m_1\big(m_2(\lambda, \cE)\big)&=-\del_\mu\big(\lambda(E^\mu-\del^\mu E)\big)
=-(\del_\mu\lambda)(E^\mu-\del^\mu E)+\lambda\,(\B E-\del_\mu E^\mu)\\
&=m_2\big(m_1(\lambda), \cE\big)+m_2\big(\lambda, m_1(\cE)\big)\;,
\end{split}    
\end{equation}
which concludes the proof of the Leibniz property \eqref{m2 formal Leibniz}.

We now turn to the proof of the homotopy associativity  relations \eqref{m2m3 Aoo}. Since the only nonzero $m_3$ involves three vector fields, most of the $m_2$ products obey strict associativity.
As we have done for the Leibniz relations, we shall proceed in order from the highest degree.

\paragraph{Degree $0$} In degree zero one can only take three gauge parameters $\lambda_1$, $\lambda_2$ and $\lambda_3$. In this case the relation is trivial to prove, due to $m_2$ being an associative pointwise product:
\begin{equation}
m_2\big(m_2(\lambda_1, \lambda_2), \lambda_3\big)=(\lambda_1\lambda_2)\lambda_3=\lambda_1(\lambda_2\lambda_3)=m_2\big(\lambda_1, m_2(\lambda_2, \lambda_3)\big)\;. \end{equation}

\paragraph{Degree $-1$} In this case the only possibility is to act on two gauge parameters $\lambda_1$, $\lambda_2$ and a field $\cA$, yielding
\begin{equation}
m_2\big(m_2(\cA, \lambda_1), \lambda_2\big)=\bpm(A^\mu\lambda_1)\lambda_2\\\del_\nu\big((A^\nu\lambda_1)\lambda_2\big)\epm =\bpm A^\mu(\lambda_1\lambda_2)\\\del_\nu\big(A^\nu(\lambda_1\lambda_2)\big)\epm  =m_2\big(\cA, m_2(\lambda_1, \lambda_2)\big)\;, 
\end{equation}
which is also strictly associative.

\paragraph{Degree $-2$} There are now two possibilities: one takes  as inputs two parameters $\lambda_1$, $\lambda_2$ and an equation of motion $\cE$, and this is associative as well:
\begin{equation}
m_2\big(m_2(\lambda_1, \lambda_2), \cE\big)=\bpm
(\lambda_1\lambda_2)(E^\mu-\del^\mu E)
\\0\epm=\bpm
\lambda_1\big(\lambda_2(E^\mu-\del^\mu E)\big)\\0\epm=m_2\big(\lambda_1, m_2(\lambda_2, \cE)\big)\;.
\end{equation}
The second possibility is to take one parameter $\lambda$ and two fields $\cA_1$, $\cA_2$, which is the first case to require a three product:
\begin{equation}
\begin{split}
m_2\big(m_2(\lambda, \cA_1),\cA_2\big)-m_2\big(\lambda, m_2(\cA_1, \cA_2)\big)&=\bpm m_2^\mu(\lambda A_1, A_2)\\0\epm-\bpm \lambda\,m_2^\mu(A_1, A_2)\\0\epm\\
&=\bpm A_1\cdot\del\lambda\,A^\mu_2+\del^\mu\lambda\,A_1\cdot A_2-2\,A_2\cdot\del\lambda\,A_1^\mu\\0\epm\\
&=m_3\big(m_1(\lambda), \cA_1, \cA_2\big)\;,
\end{split}    
\end{equation}
where we used the definition \eqref{m3}. Since the only nonzero $m_3$ has three degree $-1$ inputs, the single term appearing above is the only one required to satisfy \eqref{m2m3 Aoo}.

\paragraph{Degree $-3$} One has three cases in lowest degree: the inputs can in fact be $(\lambda_1, \lambda_2, \cN)$, $(\lambda, \cA, \cE)$ or $(\cA_1, \cA_2, \cA_3)$. The first two cases are strictly associative: 
\begin{equation}
\begin{split}
m_2\big(m_2(\lambda_1, \lambda_2), \cN\big)&=(\lambda_1\lambda_2)\,\cN=\lambda_1\,(\lambda_2\cN)=m_2\big(\lambda_1,m_2(\lambda_2, \cN)\big)\;,\\
m_2\big(m_2(\lambda, \cA), \cE\big)&=-(A^\mu\lambda)(E_\mu-\del_\mu E)=-\lambda\,\big(A^\mu(E_\mu-\del_\mu E)\big)=m_2\big(\lambda, m_2(\cA, \cE)\big)\;.
\end{split}
\end{equation}
The last relation, instead, involves the three-product and is obtained as
\begin{equation}
\begin{split}
m_2\big(m_2(\cA_1, \cA_2), \cA_3\big)-m_2\big(\cA_1, m_2(\cA_2, \cA_3)\big)&=-A_{\mu 3}\,m_2^\mu(A_1,A_2)+A_{\mu 1}\,m_2^\mu(A_2,A_3)\\
&=-\del_\mu\Big(A^\mu_1\,A_2\cdot A_3+A^\mu_3\,A_2\cdot A_1-2\,A^\mu_2\,A_1\cdot A_3\Big)\\
&=m_1\big(m_3(\cA_1, \cA_2, \cA_3)\big)\;,
\end{split}    
\end{equation}
in agreement with \eqref{m2m3 Aoo}.

In order to complete the proof that $\cK$ is a $C_\infty$-algebra without higher products, we are left to prove the consistency condition \eqref{consistent m3}. Since $m_3$ can only act on three vectors of degree $-1$, the only nontrivial relations are in degree $-2$ and $-3$.

\paragraph{Degree -2} In this case one can have three fields $\cA_i$ and one gauge parameter $\lambda$, giving rise to two possible relations:
\begin{equation}\label{m3cons1}
\begin{split}
m_2\big(m_3(\cA_1, \cA_2, \cA_3), \lambda\big)&=m_3\big(\cA_1, \cA_2, m_2(\cA_3, \lambda)\big)\;,\\
m_3\big(\cA_1, \cA_2, m_2(\cA_3, \lambda)\big)&=m_3\big(\cA_1, m_2(\cA_2, \lambda), \cA_3\big)\;.
\end{split}    
\end{equation}
The first one is easily established by computing
\begin{equation}
\begin{split}
&m_2\big(m_3(\cA_1, \cA_2, \cA_3), \lambda\big)=\lambda\bpm A_1\cdot A_2A_3^\mu+A_3\cdot A_2A_1^\mu-2\,A_1\cdot A_3A_2^\mu\\0\epm\\
&=\bpm A_1\cdot A_2(\lambda\,A_3^\mu)+(\lambda\,A_3)\cdot A_2A_1^\mu-2\,A_1\cdot (\lambda\,A_3)A_2^\mu\\0\epm=m_3\big(\cA_1, \cA_2, m_2(\cA_3, \lambda)\big)\;,
\end{split}    
\end{equation}
and the second one follows immediately from the line above.

\paragraph{Degree -3} One can only act on four fields, yielding
\begin{equation}
\begin{split}
m_2\big(m_3(\cA_1, \cA_2, \cA_3), \cA_4\big)&=-A_{4\mu}\, m_{3}^\mu(A_1, A_2, A_3)\\
&=-A_1\cdot A_2\,A_3\cdot A_4-A_3\cdot A_2\,A_1\cdot A_4+2\,A_1\cdot A_3\,A_2\cdot A_4\\
&=-A_{1\mu}\, m_{3}^\mu(A_2, A_3, A_4)=m_2\big(\cA_1, m_3(\cA_2, \cA_3, \cA_4)\big)\;,
\end{split}    
\end{equation}
which concludes the proof.

\subsection{$\mathbb{Z}_2$ Grading }\label{sec:SFT YM}

In the last part of this section we will show that both the Yang-Mills $L_\infty$-algebra $\cX$ and the $C_\infty$-algebra $\cK$ admit a further $\mathbb{Z}_2$ grading. 
This additional grading is crucial in constructing double field theory from doubling. 
To this end it is moreover convenient to write  the kinematic algebra in terms of graded basis vectors for $\cK$, which have a natural interpretation as oscillators and ghosts of an underlying $\cN=2$ particle or open string theory.

Let us start by recalling the component structure of the kinematic algebra $\cK$ (the same applies to the $L_\infty$-algebra $\cX$ upon tensoring $\cK$ with the Lie algebra $\mathfrak{g}$ and shifting the degrees accordingly). 
Specifically,  we recall that both fields and field equations are split into doublets, and we group the objects of the 
chain complex as follows
\begin{equation}
\begin{array}{cccc}
K_0&\xlongrightarrow{m_1}\;K_{-1}&\xlongrightarrow{m_1}\;K_{-2}&\xlongrightarrow{m_1}\;K_{-3}\\
\lambda&\hspace{7mm} A_{\mu} &\hspace{7mm} E&
\\
&\hspace{7mm}\varphi&\hspace{7mm}E^{\mu} &\hspace{7mm}\cN
\end{array}  \;.  
\end{equation}
This displays the decomposition of $\cK$ w.r.t.~a new $\mathbb{Z}_2$ degree with values $0,1$, which we name $c-$degree,\footnote{This degree is related to the first-quantized reparametrization ghost, as it will be explained in the following.} 
along the vertical direction above. Put differently,  
we assign $c-$degree zero to $(\lambda, A^\mu, E)$ and $c-$degree $1$ to $(\varphi, E^\mu, \cN)$, such that
\begin{equation}\label{4to6}
\begin{split}
K_{0}=K_{0}^{(0)}\;,\quad K_{-1}=K_{-1}^{(0)}\oplus K_{-1}^{(1)}\;,\quad K_{-2}=K_{-2}^{(0)}\oplus K_{-2}^{(1)}\;,\quad K_{-3}=K_{-3}^{(1)}\;.    
\end{split}    
\end{equation}
For the following discussions it will be useful to split the full complex $\cK$ according to the $c-$degree alone, writing 
\begin{equation}
\cK^{(0)}=\bigoplus_{i=-2}^0K_i^{(0)}\;,\qquad \cK^{(1)}=\bigoplus_{i=-3}^{-1}K_i^{(1)}\;. 
\end{equation}
It is important to note that $\cK^{(0)}$ and $\cK^{(1)}$ are isomorphic as vector spaces, while the $C_\infty$ degrees between the two are shifted by $-1$. 
The  vectors of $\cK^{(0)}$ and $\cK^{(1)}$ then have components
\begin{equation}\label{triplets}
\bpm\lambda\\ A^\mu\\ E\epm\in \cK^{(0)}\;,\qquad \bpm\varphi\\ E^\mu\\ \cN\epm\in \cK^{(1)}\;,   
\end{equation}
thus making the isomorphism apparent. 
The full algebra $\cK$ can thus be split as $\cK=\cK^{(0)}\oplus\cK^{(1)}$, which will be the most useful form to construct the DFT complex in the next section.

In order to simplify the subsequent treatment, in particular the doubling procedure, we will reformulate the above in a form which is more akin to the first-quantized description of the field theory. To this end, we introduce a basis for the triplet \eqref{triplets} in $\cK^{(0)}$ consisting of graded vectors
\begin{equation}\label{thetas}
\ket{\theta_M}=\Big\{\ket{\theta_+}\;,\;\ket{\theta_\mu}\;,\; \ket{\theta_-}\Big\}\;,\quad \big|\ket{\theta_M}\big|_\cK=M-1\;,    
\end{equation}
where $M=(+,\mu,-)$ count as $(+1,0,-1)$ in determining the $C_\infty$ degree.
Given the isomorphism between the components of $\cK^{(0)}$ and $\cK^{(1)}$, we shall take into account the $\mathbb{Z}_2$ split due to the $c-$degree by tensoring the above basis with a two-dimensional Grassmann algebra generated by an odd nilpotent element $c$, obeying
\begin{equation}
c^2=0\;,\quad |c|_\cK=-1\;.    
\end{equation}
This element is nothing but the reparametrization ghost of the first-quantized theory and allows one to introduce a basis $\ket{c\,\theta_M}$ for the second triplet \eqref{triplets} in $\cK^{(1)}$:
\begin{equation}\label{cthetas}
\ket{c\,\theta_M}=c\,\ket{\theta_M}\;,\quad c\,\ket{c\,\theta_M}=0\;,\quad \big|\ket{c\,\theta_M}\big|_\cK=M-2\;.  
\end{equation}
In this formulation, spacetime fields are components of intrinsic vectors of $\cK$ along the basis $\big\{\ket{\theta_M}, \ket{c\,\theta_M}\big\}$. As such, component fields are taken to have zero degree and the $C_\infty$ degree is entirely carried by the basis elements $\ket{\theta_M}$ and $\ket{c\,\theta_M}$, according to \eqref{thetas} and \eqref{cthetas}.
An arbitrary vector $u\in\cK$ can thus be expanded as
\begin{equation}\label{general decomp K}
u=\ket{\theta_M}\,u^M(x)+\ket{c\,\theta_M}\,v^M(x)\;\in\, \cK\;,   
\end{equation}
according to the decomposition $\cK=\cK^{(0)}\oplus \cK^{(1)}$.
Homogeneous vectors of the different spaces $K_i$, decomposed according to \eqref{4to6} and \eqref{triplets}, can then be written as
\begin{equation}\label{field decomposition}
\begin{split}
\Lambda&=\ket{\theta_+}\,\lambda(x)\;\in K_0\;,\quad\hspace{28mm}\cA=\ket{\theta_\mu}\,A^\mu(x)+\ket{c\,\theta_+}\,\varphi(x)\;\in K_{-1}\;,\\ \cE&=\ket{\theta_-}\,E(x)+\ket{c\,\theta_\mu}\,E^\mu(x)\;\in K_{-2}\;,\quad\cN=\ket{c\,\theta_-}\,\cN(x)\;\in K_{-3}\;,  
\end{split}    
\end{equation}
to which we will sometimes refer as
classical `string fields', gauge parameters and so on.

The same formalism can be applied at the level of the $L_\infty$-algebra $\cX$, by tensoring the graded vectors $\big\{\ket{\theta_M}, \ket{c\,\theta_M}\big\}$ with the Lie algebra generators $t_a$, yielding 
\begin{equation}\label{general decomp X}
\begin{split}
&\ket{\theta_Mt_a}=\ket{\theta_M}\otimes t_a\;,\quad \ket{c\,\theta_Mt_a}=\ket{c\,\theta_M}\otimes t_a\;,\\
&\big|\ket{\theta_Mt_a}\big|_\cX=M\;,\quad \big|\ket{c\,\theta_Mt_a}\big|_\cX=M-1\;,
\end{split}
\end{equation}
as basis elements for $\cX^{(0)}$ and $\cX^{(1)}$. However,  
for the remainder of this section we will focus on the kinematic algebra $\cK$.

In order to reformulate the differential $m_1$ and the product $m_2$ in this formalism, it is useful to introduce basis vectors for the dual space $\cK^*$. To do this, we shall first introduce another odd nilpotent element $b$ that is conjugate to $c$ and  obeys
\begin{equation}\label{b}
b^2=0\;,\quad |b|_\cK=+1\;,\quad bc+cb=1\;.    
\end{equation}
The action of $c$ and $b$ on the basis $\big\{\ket{\theta_M}, \ket{c\,\theta_M}\big\}$ realizes the $\mathbb{Z}_2$ isomorphism between $\cK^{(0)}$ and $\cK^{(1)}$ as follows:
\begin{equation}\label{bc algebra}
\begin{split}
c\,\ket{\theta_M}&=\ket{c\,\theta_M}\;,\quad c\,\ket{c\,\theta_M}=0\;,\\    
b\,\ket{\theta_M}&=0\;,\quad \hspace{8mm}b\,\ket{c\,\theta_M}=\ket{\theta_M}\;.
\end{split}    
\end{equation}
We can now introduce a basis $\big\{\bra{\theta^{M*}}, \bra{\theta^{M*}b}\big\}$ for the dual spaces $\cK^{(0)*}$ and $\cK^{(1)*}$, respectively.
The natural pairing with the basis of $\cK$ is given by
\begin{equation}\label{bra ket}
\begin{split}
\l\theta^{M*}|\theta_N\r&=\delta^M{}_N\;,\quad \l\theta^{M*}|c\,\theta_N\r=0\;,\\
\l\theta^{M*}b|\theta_N\r&=0\;,\quad \hspace{5mm}\l\theta^{M*}b|c\,\theta_N\r=\delta^M{}_N\;,
\end{split}    
\end{equation}
which fixes the degree of the dual vectors to be
$\big|\bra{\theta^{M*}}\big|_\cK=1-M$ and $\big|\bra{\theta^{M*}b}\big|_\cK=2-M$. The above pairing is consistent with the $\mathbb{Z}_2$ action of $c$ and $b$ on the dual vectors:
\begin{equation}\label{bra algebra}
\begin{split}
\bra{\theta^{M*}}\,c&=0\;,\quad \hspace{9mm}\bra{\theta^{M*}b}\,c=\bra{\theta^{M*}}\;,\\
\bra{\theta^{M*}}\,b&=\bra{\theta^{M*}b}\;,\quad \bra{\theta^{M*}b}\,b=0\;.
\end{split}    
\end{equation}
Moreover, this basis allows to write the resolution of the identity in $\cK$ as
\begin{equation}\label{1}
\mathds{1}=\ket{\theta_M}\bra{\theta^{M*}}+\ket{c\,\theta_M}\bra{\theta^{M*}b}\;.    
\end{equation}

We will now show how to reformulate the differential $m_1$ and the product $m_2$ in this formalism. We start from the differential, which is  defined as a degree $-1$ map $m_1 : \cK \rightarrow  \cK$ and which in the following we will sometimes also 
denote by $Q$ since it has the interpretation of the BRST operator of a worldline theory. We decompose  $Q$ as 
\begin{equation}\label{Q}
\begin{split}
Q \ = \ & \ket{c\,\theta_M}\bra{\theta^{M*}}\,\B \ + \ \Big(\ket{\theta_\mu}\bra{\theta^{+*}}-\ket{c\,\theta_\mu}\bra{\theta^{+*}b}\Big)\del^\mu\\
&\, +\Big(\ket{\theta_-}\bra{\theta^{\mu*}}
\ - \ \ket{c\,\theta_-}\bra{\theta^{\mu*}b}\Big)\del_\mu-\ket{\theta_-}\bra{\theta^{+*}b}\;. 
\end{split}    
\end{equation}
This notation should be understood as follows: on a general vector (\ref{general decomp K}) in $\cK$ 
the bra vectors  act in the standard fashion on ket vectors, while the 
spacetime derivatives act on the component spacetime fields, i.e., for (\ref{general decomp K}) they act on 
$u^M(x)$ and $v^M(x)$. 
So defined,  $Q$ is a map  $Q : \cK \rightarrow  \cK$, 
and 
one finds that it indeed reproduces the differential $m_1$ defined above. 
For instance, taking $u=\cA\in K_{-1}$ to be a field one has
\begin{equation}\label{Q example}
\begin{split}
Q(\cA)&=Q\Big(\ket{\theta_\nu} A^\nu+\ket{c\,\theta_+}\,\varphi\Big)
\\
&=\ket{c\,\theta_\mu}\Big(\square A^\mu-\del^\mu\varphi\Big)+\ket{\theta_-}\Big(\del\cdot A-\varphi\Big)\;\in K_{-2}\;.
\end{split}    
\end{equation}
Another important property, which can be checked by using \eqref{bc algebra} and \eqref{bra algebra}, is that
\begin{equation}\label{bBRST}
Q\,b+b\,Q=\square\, \mathds{1}\;,    
\end{equation}
which will be crucial in proving consistency of the doubling in the next section.

After discussing the realization of the differential $Q$ (or $m_1$), we now turn to the two-products \eqref{allm2s}. 
To this end it is convenient  to realize the degree zero map $m_2:\cK\times\cK\rightarrow \cK $
in terms of an element $\cM\in \cK\otimes\cK^*\otimes\cK^*$, so that 
\begin{equation}\label{m2M2}
m_2(u_1, u_2)= \cM \big(u_1\otimes u_2\big)\;,\qquad |\cM |_\cK=0\;.  
\end{equation}
The action of such an  element $\cM$ on $u_1\otimes u_2\in \cK\otimes\cK$ is defined, given 
a vector $u\in\cK$ and two dual vectors $U_1, U_2\in\cK^*$, by 
\begin{equation}\label{DualAction}
\big(u\otimes U_1\otimes U_2\big)\big(u_1\otimes u_2\big)
= \Big((-1)^{u_1U_2}\,U_1(u_1)\,U_2(u_2)\Big) u\,\in\cK \;, \end{equation}
where in the exponent we denote $u_1\equiv|u_1|_\cK$ and so on.
This is a well-defined formula for finite-dimensional vector spaces, but in our context the components are 
spacetime fields so that the algebras are infinite-dimensional and we have to be more precise about this action. 
Specifically, we define  $\cM$ by the expansion 
\begin{equation}\label{general m2}
\begin{split}
\cM&= \sum_{\alpha,\beta, \gamma=0,1}M_{\beta\gamma}^\alpha\;,\quad M^\alpha_{\beta\gamma}\;:\;\cK^{(\beta)}\otimes \cK^{(\gamma)}\;\longrightarrow\;\cK^{(\alpha)}\;,
\end{split}    
\end{equation}
and we claim  that the $m_2$ above is recovered upon setting 
\begin{equation}\label{m2 oscillators good}
\begin{split}
M^0_{00}&=\ket{\theta_+}\,\bra{\theta^{+*}}\bra{\theta^{+*}}+\ket{\theta_\mu}\Big(\bra{\theta^{\mu*}}\bra{\theta^{+*}}+\bra{\theta^{+*}}\bra{\theta^{\mu*}}\Big)\;,\\[3mm]
M^1_{00}&=\ket{c\,\theta_+}\Big(\bra{\theta^{\mu*}}\bra{\theta^{+*}}+\bra{\theta^{+*}}\bra{\theta^{\mu*}}\Big)(\del_{1\mu}+\del_{2\mu})\\[2mm]
&+\ket{c\,\theta_\mu}\Big(\bra{\theta^{\mu*}}\bra{\theta^{\nu*}}(\del_{2\nu}+2\,\del_{1\nu})-\bra{\theta^{\nu*}}\bra{\theta^{\mu*}}(\del_{1\nu}+2\,\del_{2\nu})+\bra{\theta^{\nu*}}\bra{\theta_\nu^{*}}(\del^\mu_{2}-\del^\mu_1)\Big)\\[2mm]
&-\ket{c\,\theta_\mu}\Big(\bra{\theta^{+*}}\bra{\theta^{-*}}\del_2^\mu+\bra{\theta^{-*}}\bra{\theta^{+*}}\del_1^\mu\Big)+\ket{c\,\theta_-}\Big(\bra{\theta^{\mu*}}\bra{\theta^{-*}}\del_{2\mu}+\bra{\theta^{-*}}\bra{\theta^{\mu*}}\del_{1\mu}\Big)\;,\\[3mm]
M^1_{10}&=\ket{c\,\theta_\mu}\,\bra{\theta^{\mu*}b}\bra{\theta^{+*}}+\ket{c\,\theta_-}\,\bra{\theta^{-*}b}\bra{\theta^{+*}}-\ket{c\,\theta_-}\,\bra{\theta^{\mu*}b}\bra{\theta_\mu^*}\;,\\[3mm]
M^1_{01}&=\ket{c\,\theta_\mu}\,\bra{\theta^{+*}}\bra{\theta^{\mu*}b}+\ket{c\,\theta_-}\,\bra{\theta^{+*}}\bra{\theta^{-*}b}-\ket{c\,\theta_-}\,\bra{\theta_\mu^*}\bra{\theta^{\mu*}b}\;,
\end{split}    
\end{equation}
where we omitted the tensor product symbol between the basis elements.
The action of the basis elements in here is given by (\ref{DualAction}), while the 
spacetime derivatives act on the component fields, where by $\del_{1\mu}$ and $\del_{2\mu}$ we indicate that the spacetime derivative acts on the left (respectively, right) factor of the tensor product $u_1\otimes u_2$.

As an example of how to use these formulas 
to compute the products \eqref{allm2s}, let us compute the two-product $m_2(\Lambda, \cA)$ between a gauge parameter and a field:
\begin{equation}\label{M2 example}
\begin{split}
m_2(\Lambda, \cA)&=\cM\big(\Lambda\otimes\cA\big)=
\cM\Big(\ket{\theta_+}\,\lambda\otimes\big(\ket{\theta_\mu}\,A^\mu+\ket{c\,\theta_+}\,\varphi\big)\Big)\\
&=\big(M^0_{00}+M^1_{00}\big)\big(\ket{\theta_+}\,\lambda\otimes\ket{\theta_\mu}\,A^\mu\big)+M^1_{01}\big(\ket{\theta_+}\,\lambda\otimes\ket{c\,\theta_+}\,\varphi\big)\\
&=\Big(\ket{\theta_\nu}+\ket{c\,\theta_+}(\del_{1\nu}+\del_{2\nu})\Big)\bra{\theta^{+*}}\otimes\bra{\theta^{\nu*}}\Big(\ket{\theta_+}\,\lambda\otimes\ket{\theta_\mu}\,A^\mu\Big)\\
&=\ket{\theta_\mu}\,(\lambda A^\mu)+\ket{c\,\theta_+}\,\del_\mu(\lambda A^\mu)\;,
\end{split}    
\end{equation}
where we reinstated the $\otimes$ symbol to better visualize the pairing between $\cK^*\otimes\cK^*$ and $\cK\otimes\cK$. 
We see that \eqref{M2 example} reproduces the corresponding product of \eqref{allm2s}. Finally, one may 
check that \eqref{general m2} with  components \eqref{m2 oscillators good} gives the correct symmetry property of $m_2$: 
\begin{equation}\label{cM symmetry}
\cM \big(u_1\otimes u_2\big)=(-1)^{u_1u_2}\cM\big(u_2\otimes u_1\big)\;.
\end{equation}

We are now ready to reformulate the quadratic relations of the $C_\infty$-algebra in terms of operator equations. As we have anticipated, from now on we will only be concerned with the cubic theory, whose consistency relies purely on nilpotency of the differential and the Leibniz property \eqref{m2 formal Leibniz}. Nilpotency of $m_1$ or $Q$ and 
the graded Leibniz rule \eqref{m2 formal Leibniz} can be expressed as the operator equations
\begin{equation}\label{OpLeibniz YM}
Q^2 = 0\;, \qquad Q\,\cM=\cM\,\big(Q\otimes\mathds{1}+\mathds{1}\otimes Q\big)\;,   
\end{equation}
where the right-hand side of the second equation defines the action of $Q$ on the tensor product $\cK\otimes\cK$, namely
\begin{equation}
\big(Q\otimes\mathds{1}+\mathds{1}\otimes Q\big)\big(u_1\otimes u_2\big)=\big(Q u_1\big)\otimes u_2+(-1)^{u_1}u_1\otimes \big(Qu_2\big)\;.    
\end{equation}

The Leibniz relations have been proved in section \ref{sec:Cinftyrelations}
in component form, which guarantees that \eqref{OpLeibniz YM} holds, but one can prove it directly in operator form.
For instance, one can focus on the part of \eqref{OpLeibniz YM} proportional to $\ket{\theta_\mu}\bra{\theta^{+*}}\bra{\theta^{+*}}$ and compute
\begin{equation}
Q\,\cM\big\rvert^{++}_\mu=\Big(\ket{\theta_\mu}\bra{\theta^{+*}}\del^\mu\Big)\ket{\theta_+}\bra{\theta^{+*}}\bra{\theta^{+*}}=\ket{\theta_\mu}\bra{\theta^{+*}}\bra{\theta^{+*}}(\del^\mu_1+\del^\mu_2)\;,    
\end{equation}
where we wrote the total derivative in $Q\,\cM$ as $\del^\mu_1+\del^\mu_2$ acting on $\cK\otimes\cK$. 
Similarly, the corresponding part of the right-hand side of \eqref{OpLeibniz YM} yields
\begin{equation}
\begin{split}
&\cM\,\big(Q\otimes\mathds{1}+\mathds{1}\otimes Q\big)\big\rvert^{++}_\mu\\
&=
\ket{\theta_\mu}\bra{\theta^{\mu*}}\otimes\bra{\theta^{+*}}\Big(\ket{\theta_\nu}\bra{\theta^{+*}}\del_1^\nu\otimes\mathds{1}\Big)+\ket{\theta_\mu}\bra{\theta^{+*}}\otimes\bra{\theta^{\mu*}}\Big(\mathds{1}\otimes\ket{\theta_\nu}\bra{\theta^{+*}}\del_2^\nu\Big)\\
&=\ket{\theta_\mu}\bra{\theta^{+*}}\bra{\theta^{+*}}(\del^\mu_1+\del^\mu_2)\;.
\end{split}    
\end{equation}
This proves the 
$\ket{\theta_\mu}\bra{\theta^{+*}}\bra{\theta^{+*}}$ `component' of \eqref{OpLeibniz YM}, and
all the other components can be proved in the same way.
In the next section we will use two copies of the vector space $\cK$, each one endowed with  its $Q$ and $\cM$ operators, to construct double field theory at cubic order. Consistency of the latter will be a direct consequence of the fundamental relations
(\ref{OpLeibniz YM}).

\section{Cubic Double Field Theory from Doubling}\label{sec:DFT}

We start this section by describing the $L_\infty$ complex $\cV$ for double field theory (DFT), as given in 
the original formulation of Hull and Zwiebach \cite{Hull:2009mi}, and 
show that $\cV$ is a subspace of the tensor product of two copies of the $C_\infty$ complex $\cK$ of Yang-Mills theory. 
We then define the DFT differential $\Q$ and two-bracket $B_2$ in terms of two copies of the Yang-Mills operators $Q$ and 
$\cM$ and prove that $\mathds{Q}$ is nilpotent and acts as a derivation of $B_2$. This implies consistency of DFT 
at cubic order. We apply these results  by giving the explicit expressions for the deformed gauge transformations and
the cubic action, which yields  significant simplifications of the results in \cite{Hull:2009mi}.

\subsection{The Double Field Theory Complex}

Double field theory is defined on a doubled spacetime, which we take to have coordinates $(x^\mu, \bar x^{\bar{\mu}})$. These correspond to the left and right-moving
parts of the closed string center of mass, while the coordinates more often used in the literature, $(X^\mu, \widetilde X_\mu)$, correspond to the standard and dual frames with respect to $T-$duality. 
In addition, all fields and parameters in DFT are subject to the weak constraint, corresponding to the level matching constraint in string theory. In the coordinate system $(x^\mu, \bar x^{\bar \mu})$ this is expressed as \cite{Hull:2009mi}
\begin{equation}\label{weak}
\Delta=\tfrac12\big(\B-\Bb\big)=0\;,   
\end{equation}
where $\B=\eta^{\mu\nu}\partial_{\mu}\partial_{\nu}$ and 
$\Bb=\eta^{\bar{\mu}\bar{\nu}}\bar{\partial}_{\bar{\mu}}\bar{\partial}_{\bar{\nu}}$ 
are defined in terms of two copies of the Minkowski metric $\eta$ and the derivatives\footnote{The metric signature plays no role in this construction, 
and so we can take $\eta$ to be Lorentzian even though in \cite{Hull:2009mi} the metric is the flat Euclidean metric on a torus.}  
\begin{equation}
\del_\mu=\frac{\del}{\del x^\mu}\;,\quad \bdel_{\bar \mu}=\frac{\del}{\del \bar x^{\bar \mu}}\;. 
\end{equation}
The standard supergravity solution of the weak constraint is to identify the two sets of coordinates, declaring 
$\del_\mu=\bdel_{\bar \mu}$ on all fields and parameters, which eliminates the dependence on the dual coordinates $\widetilde X_\mu$.

We will now proceed to describe the field content of double field theory. In the formulation of \cite{Hull:2009mi}, which arises from closed string field theory, the fields consist of a tensor $e_{\mu\bar{\nu}}$ with no symmetry between the indices, a couple of scalar fields $e$ and $\bar e$, and a couple of vectors $f_\mu$ and $\bar f_{\bar{\mu}}$. The tensor $e_{\mu\bar{\nu}}$ contains the graviton and the $B-$field as its symmetric and antisymmetric parts, respectively. The two scalars account for the dilaton $d=\frac12 (e-\bar e)$ and a pure gauge scalar $\rho=\frac12 (e+\bar e)$, while the vectors $f_\mu$ and $\bar f_{\bar{\mu}}$ are auxiliary fields, similar to the scalar $\varphi$ of Yang-Mills.
Associated to the tensor $e_{\mu\bar{\nu}}$ one has two independent vector gauge symmetries, with parameters $\lambda_\mu$ and $\bar\lambda_{\bar{\mu}}$, which are related to metric diffeomorphisms and $B-$field gauge transformations. Due to the presence of auxiliaries and of the pure gauge scalar $\rho$, we also have a scalar St\"uckelberg symmetry with parameter $\eta$. Finally, the presence of a two-form inside $e_{\mu\bar{\nu}}$ implies that the gauge transformations are reducible, with a gauge-for-gauge parameter $\chi$.

In $L_\infty$ language, these  fields, gauge parameters, field equations and so on are elements of a graded vector space $\cV$ which decomposes, in the $b-$picture, as the direct sum
\begin{equation}
\cV=\bigoplus_{i=-3}^{+2}V_i \;.   
\end{equation}
In non-negative degree it contains the gauge-for-gauge parameter $\chi$ and the multiplets of gauge parameters $\Lambda$ and fields $\psi$:
\begin{equation}\label{Zfields1}
\chi\in V_2\;,\quad
\Lambda=\bpm\lambda_\mu\;,\;\bar\lambda_{\bar \mu}\\
\eta\epm \in V_1\;,\quad 
\psi =\bpm e_{\mu\bar{\nu}}\\
f_\mu\;,\;\bar f_{\bar{\mu}}\\
e\;,\;\bar e\epm\in V_0\;. 
\end{equation}
The negative degree spaces consist of the field equations $\cF$ (dual to fields), Noether identities $\cN$ and the Noether-for-Noether identity 
$\cR$ dual to reducibility:
\begin{equation}\label{Zfields2}
\cF=\bpm F_{\mu\bar{\nu}}\\
F_\mu\;,\;\bar F_{\bar \mu}\\
F\;,\;\bar F\epm\in V_{-1}\;,\quad \cN=\bpm N_\mu\;,\;\bar N_{\bar \mu}\\
N\epm \in V_{-2}\;,\quad\cR\in V_{-3}\;.  
\end{equation}
A nilpotent differential $b_1$, which will be discussed in the following, maps between these spaces, making $\cV$ into a chain complex:
\begin{equation}
\begin{array}{cccccc}
V_2&\xlongrightarrow{b_1}\;V_1&\xlongrightarrow{b_1}\;V_0&\xlongrightarrow{b_1}\;V_{-1}&\xlongrightarrow{b_1}\;V_{-2}&\xlongrightarrow{b_1}\;V_{-3}\\
\chi&\hspace{7mm}\Lambda&\hspace{7mm}\psi&\hspace{7mm}\cF&\hspace{7mm}\cN&\hspace{7mm}\cR
\end{array}  \;.  
\end{equation}

We conclude this presentation by introducing a further $\mathbb{Z}_2$ grading of $\cV$, which we name $c^+-$degree and
which  is analogous to the one introduced for $\cK$ in Yang-Mills. The vector space $\cV$ can thus be split into two components according to their $c^+-$degree:
\begin{equation}
\cV=\cV^{(0)}\oplus \cV^{(1)}\;,    
\end{equation}
where we assign $c^+-$degree zero and one to the component fields as
\begin{equation}\label{Z0Z1 components}
\bpm \chi\\
\lambda_\mu\;,\;\bar\lambda_{\bar \mu}\\
e\;,\;e_{\mu\bar{\nu}}\;,\;\bar e\\
\bar F_{\bar \mu}\;,\; F_\mu\\
N
\epm\in\cV^{(0)}\;,\quad
\bpm \eta\\
f_\mu\;,\;\bar f_{\bar \mu}\\
F\;,\;F_{\mu\bar{\nu}}\;,\;\bar F\\
\bar N_{\bar \mu}\;,\; N_\mu\\
\cR
\epm\in\cV^{(1)}\;.
\end{equation}
Notice that fields in \eqref{Z0Z1 components} are organized vertically by decreasing $L_\infty$ degree and, as in the Yang-Mills case, one can see that $\cV^{(0)}$ and $\cV^{(1)}$ are isomorphic as vector spaces, with $L_\infty$ degrees shifted by $-1$ between the two.

We are now in the position to show that the complex $\cV$ of double field theory is a subspace of the tensor product $\cK\otimes\overline\cK$ of two $C_\infty$-algebras of Yang-Mills. To this end, let us consider two copies $\cK$ and $\overline\cK$ of the Yang-Mills kinematic algebras. As we have seen in section \ref{sec:SFT YM}, these split according to their respective $c-$degrees as 
\begin{equation}
\cK=\cK^{(0)}\oplus\cK^{(1)}\;,\quad  \overline\cK=\overline\cK^{(0)}\oplus\overline\cK^{(1)}\;.  
\end{equation} 
From the decomposition \eqref{triplets} of Yang-Mills theory, one can see that the 
$L_\infty$ vector space of DFT  given by the direct sum $\cV=\cV^{(0)}\oplus\cV^{(1)}$
can be accommodated in the tensor products 
\begin{equation}\label{Z0Z1 tensor products}
\cV^{(0)}=\cK^{(0)}\otimes\overline\cK^{(0)}\;,\quad \cV^{(1)}=\big(\cK^{(0)}\otimes\overline\cK^{(1)}\big)\oplus\big(\cK^{(1)}\otimes\overline\cK^{(0)}\big)\;.
\end{equation}
In particular, we take the tensor products of spaces of functions of $x$ and functions of $\bar{x}$ to give 
functions of $(x,\bar{x})$, as in (\ref{tensoringfunctions}).

The tensor product requires a degree shift, which we describe in the following. 
To this end it is convenient to introduce  a basis.  
We take the basis vectors for the two copies $\cK$ and $\overline\cK$ to be $\big\{\ket{\theta_M}, \ket{c\,\theta_M} \big\}$ and $\big\{\ket{\bar\theta_{\bar M}}, \ket{\bar c\,\bar\theta_{\bar M}} \big\}$, respectively, according to our discussion in section \ref{sec:SFT YM}.
Similarly, we introduce two copies of the dual vectors: $\big\{\bra{\theta^{M*}}, \bra{\theta^{M*}b} \big\}$ and $\big\{\bra{\bar\theta^{\bar{M}*}}, \bra{\bar\theta^{\bar{M}*}\bar b} \big\}$, each obeying the relations \eqref{bra ket} and \eqref{bra algebra} with the respective $(b,c)$ and $(\bar b, \bar c)$ operators. We introduce now the linear combinations
\begin{equation}\label{bc+-}
c^\pm:=c\pm\bar c:=c\otimes\mathds{1}\pm\mathds{1}\otimes\bar c\;,\qquad b^\pm:=\tfrac12(b\pm\bar b):=\tfrac12 (b\otimes\mathds{1}\pm\mathds{1}\otimes\bar b)\;, 
\end{equation}
defined as above on $\cK\otimes\overline{\cK}$.
They obey
\begin{equation}\label{bc+- algebra}
\big(c^\pm\big)^2=0\;,\quad \big(b^\pm\big)^2=0\;,\quad b^\pm c^\pm+c^\pm b^\pm=1\;,\quad b^\pm c^\mp+c^\mp b^\pm=0\;,   
\end{equation}
and allow us to write down a basis for $\cV^{(0)}$ and $\cV^{(1)}$ as  
\begin{equation}\label{Z basis}
\begin{split}
\ket{\theta_M\bar\theta_{\bar N}}&=\ket{\theta_M}\otimes\ket{\bar\theta_{\bar N}}\;,\\
\ket{c^+\theta_M\bar\theta_{\bar N}}&=\ket{c\,\theta_M}\otimes\ket{\bar\theta_{\bar N}}+(-1)^{M-1}\ket{\theta_M}\otimes\ket{\bar c\,\bar\theta_{\bar N}}\;,   
\end{split}
\end{equation}
in terms of the natural basis $\big\{\ket{\theta_M}, \ket{c\,\theta_M} \big\}\otimes\big\{\ket{\bar\theta_{\bar{N}}}, \ket{\bar c\,\bar\theta_{\bar N}} \big\}$ of $\cK\otimes\overline{\cK}$. The isomorphism between $\cV^{(0)}$ and $\cV^{(1)}$ is given by the action of $(b^+, c^+)$ as
\begin{equation}
\begin{split}
c^+\,\ket{\theta_M\bar\theta_{\bar N}}&=\ket{c^+\theta_M\bar\theta_{\bar N}}\;,\quad c^+\,\ket{c^+\theta_M\bar\theta_{\bar N}}
=0\;,\\    
b^+\,\ket{\theta_M\bar\theta_{\bar N}}&=0\;,\quad \hspace{15mm}b^+\,\ket{c^+\theta_M\bar\theta_{\bar N}}=\ket{\theta_M\bar\theta_{\bar N}}\;.    
\end{split}    
\end{equation}
An arbitrary element $\Psi\in\cV$ can thus be expanded as
\begin{equation}\label{psichi DFT}
\Psi=\ket{\theta_M\bar\theta_{\bar N}}\,\phi^{M\bar{N}}
+ \ket{c^+\theta_M\bar\theta_{\bar N}}\,\chi^{M\bar{N}}\;,   
\end{equation}
where both $\phi$ and $\chi$ depend on doubled coordinates $(x,\bar{x})$. 
In order to match the $L_\infty$ degrees of the $\cV$ complex, one has to shift by two the $C_\infty$ degrees of the tensor product: given $u\in\cK$ and $\bar u\in\overline{\cK}$, one defines
\begin{equation}\label{suspension}
\Psi=u\otimes\bar u\in\cV\;,\quad |\Psi|_\cV=|u|_\cK+|\bar u|_{\overline{\cK}}+2\;,    
\end{equation}
which implies the following assignment for the basis vectors and the $(b^\pm, c^\pm)$ operators:
\begin{equation}
\big|\ket{\theta_M\bar\theta_{\bar N}}\big|_\cV=M+N\;,\quad\big|\ket{c^+\theta_M\bar\theta_{\bar N}}\big|_\cV=M+N-1\;,\quad \big|c^\pm\big|_\cV=-1\;,\quad \big|b^\pm\big|_\cV=+1\;.
\end{equation}
Using this basis one can rewrite the $L_{\infty}$ elements  \eqref{Zfields1} and \eqref{Zfields2} of homogeneous degree as
\begin{equation}\label{Z a lot}
\begin{split}
\chi&=\ket{\theta_+\bar\theta_+}\,\chi\in V_2\;,\\
\Lambda&=\ket{\theta_+\bar\theta_{\bar \mu}}\,\bar\lambda^{\bar \mu}-\ket{\theta_\mu\bar\theta_+}\,\lambda^\mu-2\,\ket{c^+\theta_+\bar\theta_+}\,\eta\in V_1\;,\\
\psi
&=\ket{\theta_\mu\bar\theta_{\bar \nu}}\,e^{\mu\bar{\nu}}+2\,\ket{\theta_+\bar\theta_-}\,\bar e+2\,\ket{\theta_-\bar\theta_+}\, e+2\,\ket{c^+\theta_+\bar\theta_{\bar\mu}}\,\bar f^{\bar \mu}+2\,\ket{c^+\theta_\mu\bar\theta_+}\, f^\mu\in V_0\;,\\
\cF&=\ket{c^+\theta_\mu\bar\theta_{\bar \nu}}\,F^{\mu\bar{\nu}}+\ket{c^+\theta_+\bar\theta_-}\,\bar F+\ket{c^+\theta_-\bar\theta_+}\, F+\ket{\theta_\mu\bar\theta_-}\, F^\mu+\ket{\theta_-\bar\theta_{\bar \mu}}\,\bar F^{\bar \mu}\in V_{-1}\;,\\
\cN&=2\,\ket{c^+\theta_-\bar\theta_{\bar \mu}}\,\bar N^{\bar \mu}-2\,\ket{c^+\theta_\mu\bar\theta_-}\,N^\mu-\ket{\theta_-\bar\theta_-}\,N\in V_{-2}\;,\\ \cR&=-\ket{c^+\theta_-\bar\theta_-}\,\cR\in V_{-3}\;,
\end{split}    
\end{equation}
where the normalizations have been chosen to match the ones of \cite{Hull:2009mi}.

We conclude this subsection by giving an alternative and more useful characterization 
 of the vector space $\cV$ as a subspace of the full tensor product $\cK\otimes\overline\cK$. 
 Taking linear combinations of the natural basis  vectors $\ket{\theta_M}\otimes\ket{\bar\theta_{\bar N}}$, $\ket{c\,\theta_M}\otimes\ket{\bar\theta_{\bar N}}$, $\ket{\theta_M}\otimes\ket{\bar c\,\bar\theta_{\bar N}}$ and 
 $\ket{c\,\theta_M}\otimes\ket{\bar c\,\bar\theta_{\bar N}}$ of $\cK\otimes\overline\cK$ we  define the new  basis
\begin{equation}\label{KtildeK basis}
\ket{\theta_M\bar\theta_{\bar N}}=\ket{\theta_M}\otimes\ket{\bar\theta_{\bar N}}\;,\quad \ket{c^\pm\theta_M\bar\theta_{\bar N}}=c^\pm\ket{\theta_M\bar\theta_{\bar N}}\;,\quad \ket{c^-c^+\theta_M\bar\theta_{\bar N}}=c^-c^+\ket{\theta_M\bar\theta_{\bar N}}\;,
\end{equation}
where the action of $(b^\pm, c^\pm)$ follows with the algebra \eqref{bc+- algebra}. This 
shows that vectors of $\cV$ that have components only along $\ket{\theta_M\bar\theta_{\bar N}}$ and $\ket{c^+\theta_M\bar\theta_{\bar N}}$ are the vectors $\Psi$ of $\cK\otimes\overline{\cK}$ obeying $b^-\Psi=0$. This is analogous to the constraint usually imposed in closed string field theory \cite{Zwiebach:1992ie}. Recalling that all fields in $\cV$ are subject to the weak constraint \eqref{weak}, the DFT complex $\cV$ can be  defined as the subspace 
\begin{equation}\label{Z in KtildeK}
\cV=\left\{\Psi\in\cK\otimes\overline\cK\;|\;  b^-\Psi=0\,,\;\Delta\Psi=0\right\}\;.
\end{equation}

\subsection{$L_{\infty}$-algebra of Double Field Theory to Cubic Order}

We will now construct cubic double field theory by identifying its differential $\mathds{Q}$ and two-bracket $B_2$. It will be shown that these obey the quadratic relations
\begin{equation}\label{quadratic relations DFT}
\begin{split}
\mathds{Q}^2(\Psi)&=0\;,\quad\forall\,\Psi\in \cV\;,\\    
\mathds{Q}\big(B_2(\Psi_1,\Psi_2)\big)+B_2\big(\mathds{Q}(\Psi_1),\Psi_2\big)+(-1)^{|\Psi_1|}B_2\big(\Psi_1, \mathds{Q}(\Psi_2)\big)&=0\;,\quad\forall\,\Psi_1, \Psi_2\in \cV\;,
\end{split}    
\end{equation}
which ensure consistency up to cubic order.
In order to prove nilpotency of $\mathds{Q}$ and the Leibniz property, it will be instrumental to consider $\cV$ as the subspace \eqref{Z in KtildeK} and work on $\cK\otimes\overline{\cK}$.
We start from the differential, which defines the free theory. Let us consider two copies $Q$ and $\overline{Q}$ of the Yang-Mills differential \eqref{Q}, obeying
\begin{equation}\label{QQbar}
\begin{split}
Q\;:\;\cK\;\longrightarrow\;\cK\;,\quad|Q|_\cK=-1\;,\quad Q^2=0\;,\\    
\overline 
Q\;:\;\overline\cK\;\longrightarrow\;\overline\cK\;,\quad|\overline Q|_{\overline{\cK}}=-1\;,\quad \overline Q^2=0\;,
\end{split}    
\end{equation}
and define their sum $\big(Q+\overline{Q}\big)$, acting on $\cK\otimes\overline{\cK}$ as
\begin{equation}\label{Q+Qbar}
\begin{split}
Q+\overline{Q}&=Q\otimes\mathds{1}+\mathds{1}\otimes\overline{Q}\;,\\
\big(Q+\overline{Q}\big)\big(u\otimes\bar u\big)&=\big(Q\,u\big)\otimes\bar u+(-1)^{u}u\otimes\big(\overline{Q}\,\bar u\big)\;.
\end{split}    
\end{equation}
As defined, this operator is not a map from $\cV$ to $\cV$, but rather from the full $\cK\otimes\overline{\cK}$ to itself. We will now prove, however, that $Q+\overline{Q}$ is well-defined on $\cV$, meaning that
\begin{equation}\label{Q+Q well defined}
\forall\,\Psi\in\cV\subset\cK\otimes\overline{\cK}\quad\Longrightarrow\quad \big(Q+\overline{Q}\big)(\Psi)\in\cV  \;.  
\end{equation}
The first requirement for this is that $Q+\overline{Q}$ preserves the weak constraint. This is trivial, since $\Delta$ in \eqref{weak} commutes with both $Q$ and $\overline{Q}$, as is evident from (\ref{Q}).  The second requirement is to preserve the space $\ker(b^-)$, according to \eqref{Z in KtildeK}. In order to see this, let us compute $\big(Q+\overline{Q}\big)b^-$:
\begin{equation}
\begin{split}
\big(Q+\overline{Q}\big)b^-&=\tfrac12\big(Q\otimes\mathds{1}+\mathds{1}\otimes\overline{Q}\big)\big(b\otimes\mathds{1}-\mathds{1}\otimes\bar b\big)\\
&=\tfrac12\,\Big[\big(Q\,b\big)\otimes\mathds{1}-Q\otimes\bar b-b\otimes\overline{Q}-\mathds{1}\otimes\big(\overline{Q}\,\bar b\big)\Big]\;,
\end{split}    
\end{equation}
where signs are determined by all operators having odd degree.
One can compute $b^-\big(Q+\overline{Q}\big)$ in the same way and obtain the anticommutator
\begin{equation}\label{QbarQ anticom b-}
\begin{split}
\big(Q+\overline{Q}\big)b^-+b^-\big(Q+\overline{Q}\big)&=\tfrac12\,\Big[\big(Q\,b+b\,Q\big)\otimes\mathds{1}-\mathds{1}\otimes\big(\overline{Q}\,\bar b+\bar b\,\overline{Q}\big)\Big]\\
&=\tfrac12\,\big(\B-\Bb\big)\,\mathds{1}=\Delta\mathds{1}\;,
\end{split}
\end{equation}
where we used \eqref{bBRST}.
It is now easy to see that $Q+\overline{Q}$ is well-defined on $\cV$:
\begin{equation}
\begin{split}
b^-\Psi&=0\;,\quad\Delta\Psi=0\;\longrightarrow\\
b^-\big(Q+\overline{Q}\big)(\Psi)&=-\big(Q+\overline{Q}\big)b^-\Psi+\Delta\Psi=0\;.
\end{split}    
\end{equation}
Since $Q+\overline{Q}$ preserves the $\cV$ subspace of $\cK\otimes\overline{\cK}$, we shall define the DFT differential $\mathds{Q}$ as the restriction
\begin{equation}
\mathds{Q}=\big(Q+\overline{Q}\big)\big\rvert_\cV \;,\quad{\rm s.t.}\quad \forall\,\Psi\in\cV\;,\quad \mathds{Q}(\Psi)=\big(Q+\overline{Q}\big)(\Psi)\;. \end{equation}
Nilpotency or $\mathds{Q}^2=0$ then follows immediately from $\big(Q+\overline{Q}\big)^2=0$, 
which in turn follows from \eqref{Q+Qbar} together with $Q^2=\overline{Q}^2=0$ and $Q$ and $\overline{Q}$ being  odd. 
According to the degree assignment \eqref{suspension} and the definition \eqref{Q+Qbar},  the $L_\infty$ degree is given by $\big|\Q\big|_\cV=\big|Q+\overline{Q}\big|_\cV=-1$, as it should.

As a concrete example of the action of $\Q$, let us compute the gauge transformation of the tensor field $e^{\mu\bar{\nu}}$ 
from the general formula $\delta\psi=\Q(\Lambda)$.  Given the basis decomposition \eqref{Z a lot}, this is obtained as the $\ket{\theta_\mu\bar\theta_{\bar \nu}}$ component of $\Q(\Lambda)$. Using the form \eqref{Q} of $Q$ (and the same for $\overline{Q}$), one has
\begin{equation}
\begin{split}
\delta e^{\mu\bar{\nu}}&=\bra{\bar\theta^{\bar\nu*}}\bra{\theta^{\mu*}}\Q(\Lambda)=\bra{\bar\theta^{\bar\nu*}}\bra{\theta^{\mu*}}\Big\{\Big(\ket{\theta_\rho}\bra{\theta^{+*}}\del^\rho\otimes\mathds{1}+\mathds{1}\otimes\ket{\bar\theta_{\bar\rho}}\bra{\bar\theta^{+*}}\bdel^{\bar \rho}\Big)\\
&\hspace{53mm}\Big(\ket{\theta_+\bar\theta_{\bar \sigma}}\,\bar\lambda^{\bar \sigma}-\ket{\theta_\sigma\bar\theta_+}\,\lambda^\sigma-2\,\ket{c^+\theta_+\bar\theta_+}\,\eta\Big)\Big\}\\
&=\Big(\bra{\theta^{+*}}\del^\mu\otimes\bra{\bar\theta^{\bar\nu*}}-\bra{\theta^{\mu*}}\otimes\bra{\bar\theta^{+*}}\bdel^{\bar \nu}\Big)\Big(\ket{\theta_+\bar\theta_{\bar \sigma}}\,\bar\lambda^{\bar \sigma}-\ket{\theta_\sigma\bar\theta_+}\,\lambda^\sigma-2\,\ket{c^+\theta_+\bar\theta_+}\,\eta\Big)\\
&=\del^\mu\bar\lambda^{\bar \nu}+\bdel^{\bar \nu}\lambda^\mu\;,
\end{split}    
\end{equation}
where in the first step we excluded all components of $\Q$ which have zero overlap with $\bra{\bar\theta^{\bar\nu*}}\bra{\theta^{\mu*}}$. The rest of the gauge transformations can be computed in the same way, yielding the component expressions
\begin{equation}\label{gauge free DFT}
\begin{split}
\delta e_{\mu\bar{\nu}} &= \del_\mu\bar\lambda_{\bar \nu}+\bdel_{\bar \nu}\lambda_\mu\;,\\
\delta f_\mu&=-\tfrac12\,\B\lambda_\mu+\del_\mu\eta\;,\qquad 
\delta\bar f_{\bar \mu}=\tfrac12\,\B\bar\lambda_{\bar \mu}+\bdel_{\bar \mu}\eta\;,\\
\delta e&=-\tfrac12\,\del\cdot\lambda+\eta\;,\qquad \hspace{6mm}\delta \bar e=\tfrac12\,\bar{\del}\cdot\bar\lambda+\eta\;,
\end{split}    
\end{equation}
where we used $\B=\Bb$ on weakly constrained fields.
The linearized reducibility is encoded in parameters of the form $\Lambda=\mathds{Q}(\chi)$, i.e., 
\begin{equation}
\lambda_\mu=-\del_\mu\chi\;,\qquad \bar\lambda_{\bar\mu}=\bdel_{\bar \mu}\chi\;,\qquad \eta=-\tfrac12\,\B\chi\;,
\end{equation}
which  generate no gauge transformations at this order.
The free field equations are given by $\mathds{Q}(\psi)=0$, with the string field $\psi$ as in \eqref{Z a lot}. Expanding $\Q(\psi)$ in components one finds: 
\begin{equation}
\begin{split}
&\B\, e_{\mu\bar{\nu}}+2\,\bdel_{\bar \nu} f_\mu-2\,\del_\mu\bar f_{\bar \nu}=0    \;,\\
&\B\, e-\del\cdot f=0\;,\qquad \B\, \bar e-\bdel\cdot\bar  f=0\;,\\
&\del_\mu\bar e-\tfrac12\,\bdel^{\bar \rho} e_{\mu\bar{\rho}}-f_\mu=0\;,\qquad\bdel_{\bar{\mu}} e +\tfrac12\,\del^\rho 
e_{\rho\bar{\mu}}-\bar f_{\bar \mu}=0\;,
\end{split}    
\end{equation}
in agreement with \cite{Hull:2009mi}. Finally, given the field equations $\cF$, the Noether identities are obtained by $\cN=\mathds{Q}(\cF)$ and, analogously, $\cR=\mathds{Q}(\cN)$ is the Noether-for-Noether identity.

Having determined the DFT differential, the next goal is to find the two-bracket $B_2$ starting from two copies of the kinematic products $\cM$ and $\overline \cM$. The main idea comes from the fact that the separate copies of $\cM$ obey the Leibniz relations in the operator form \eqref{OpLeibniz YM}, as discussed in section \ref{sec:SFT YM}.
In this formalism, the two-bracket of double field theory can similarly be written as an element $\cB\in\cV\otimes\cV^*\otimes\cV^*$ acting on the tensor product of two vectors in $\cV$:
\begin{equation}
B_2\big(\Psi_1, \Psi_2\big)=\cB\big(\Psi_1\otimes\Psi_2\big)\;,\quad\big|\cB\big|_\cV=-1\;,   
\end{equation}
where the action on the tensor product $\cV\otimes\cV$ is defined analogously to \eqref{DualAction}. The Leibniz relation in \eqref{quadratic relations DFT} can be expressed as
\begin{equation}\label{B2 leibniz operator}
\mathds{Q}\,\cB+\cB\,\big(\Q\otimes\mathds{1}+\mathds{1}\otimes \Q\big)=0\;,  
\end{equation}
where the second term defines the action of $\Q$ on the tensor product $\cV\otimes\cV$, namely
\begin{equation}
\big(\Q\otimes\mathds{1}+\mathds{1}\otimes \Q\big)\big(\Psi_1\otimes\Psi_2\big)=(\mathds{Q}\Psi_1)\otimes\Psi_2+(-1)^{\Psi_1}\Psi_1\otimes(\mathds{Q}\Psi_2)\;.    
\end{equation}
Let us now discuss the ansatz for the two-bracket. The naive guess $\cB\sim \cM\otimes\overline\cM$ cannot work for two reasons. First of all, this naive ansatz does not respect the weak constraint, i.e.~generally $\big(\cM\otimes\overline{\cM}\big)\big(\Psi_1\otimes\Psi_2\big)\notin\cV$ for $\Psi_1, \Psi_2\in\cV$. Moreover, the degree is wrong: the operator $\cM$ carries degree $|\cM|_\cK=0$ in $\cK$, same for $\overline{\cM}$ in $\overline{\cK}$. According to the degree shift \eqref{suspension}, one has $\big|\cM\otimes\overline \cM\big|_\cV=-2$, while $\big|\cB\big|_\cV=-1$. The simplest ansatz which maintains the structure of a tensor product and solves both issues is given by
\begin{equation}\label{doubling ansatz}
\cB=-\tfrac12\,\cP_\Delta\,b^-\cM\otimes\overline\cM\;,
\end{equation}
where the overall normalization $-\tfrac12$ can be chosen at will and has been fixed to make contact with the literature.
In \eqref{doubling ansatz} we have introduced the projector $\cP_\Delta$ to $\ker(\Delta)$, satisfying 
\begin{equation}
\cP_\Delta^2=\cP_\Delta\;,\quad\Delta\,\cP_\Delta=0\;,\quad\cP_\Delta\,\Delta=0\;.    
\end{equation}
The ansatz \eqref{doubling ansatz} automatically respects both the weak and the algebraic constraints, in that $b^-\,\cB=\Delta\,\cB=0$ thanks to the explicit projection.
Given two vectors $\Psi_1, \Psi_2\in\cV$ of the form $\Psi_i=u_i\otimes\bar u_i$, with $u_i\in\cK$ and $\bar u_i\in\overline{\cK}$, the action of $\cB$ is defined by
\begin{equation}\label{B action}
\begin{split}
\cB\big(\Psi_1\otimes\Psi_2\big)&= -\tfrac12\,\cP_\Delta\,b^-\big(\cM\otimes\overline\cM\big)\Big(\big(u_1\otimes\bar u_1\big)\otimes \big(u_2\otimes\bar u_2\big)\Big)\\
&=-\tfrac12\,(-1)^{\bar u_1u_2}\,\cP_\Delta\,b^-\cM\big(u_1\otimes u_2\big)\otimes\overline{\cM}\big(\bar u_1\otimes \bar u_2\big)\;,
\end{split}    
\end{equation}
where we denoted $\bar u_1=|\bar u_1|_{\overline{\cK}}$, $u_2=| u_2|_{{\cK}}$ in the exponent, and we recall that the action of $b^-$ on $\cK\otimes\overline{\cK}$ is given by \eqref{bc+-}. 

We start checking the consistency of the construction by verifying the symmetry property of the two-bracket $\cB$. Taking two elements  $\Psi_i=u_i\otimes\bar u_i$ with $i=1,2$ we  compute
\begin{equation}\label{L2 anti}
\begin{split}
\cB\big(\Psi_1\otimes\Psi_2\big)&=-\tfrac12\,(-1)^{\bar u_1u_2}\,\cP_\Delta\,b^-\cM\big(u_1\otimes u_2\big)\otimes\overline{\cM}\big(\bar u_1\otimes \bar u_2\big)\\
&=-\tfrac12\,(-1)^{\bar u_1u_2+u_1u_2+\bar u_1\bar u_2}\,\cP_\Delta\,b^-\cM\big(u_2\otimes u_1\big)\otimes\overline{\cM}\big(\bar u_2\otimes \bar u_1\big)\\
&=(-1)^{\bar u_1u_2+u_1u_2+\bar u_1\bar u_2+\bar u_2u_1}\,\cB\big(\Psi_2\otimes\Psi_1\big)\\
&=(-1)^{\Psi_1\Psi_2}\,\cB\big(\Psi_2\otimes\Psi_1\big)\;,
\end{split}    
\end{equation}
which is the expected graded symmetry of the two-bracket in the $b-$picture. In order to obtain \eqref{L2 anti} we have used the symmetry property \eqref{cM symmetry} and \eqref{suspension}, namely that $|\Psi|_\cV=|u|_\cK+|\bar u|_{\overline\cK}+2$ for $\Psi=u\otimes\bar u$.
We can now prove that the two-bracket $\cB$ satisfies the Leibniz relation in the form \eqref{B2 leibniz operator}. To do so, it is important to recall that $\Q=Q+\overline{Q}$ when acting on $\cV$, as well as the identity \eqref{QbarQ anticom b-}, which allow us to write
\begin{equation}\label{useful id}
\mathds{Q}\,\cP_\Delta\,b^-=\big(Q+\overline Q\big)\cP_\Delta\,b^-=-\cP_\Delta\,b^-\big(Q+\overline Q\big)\;,    
\end{equation}
where the first equality comes from the fact that $\cP_\Delta\,b^-\,\Psi\in\cV$ for any $\Psi\in\cK\otimes\overline{\cK}$, while the second one holds under the projector $\cP_\Delta$ thanks to \eqref{QbarQ anticom b-}. The last ingredient to prove the Leibniz property is that $Q$ and $\overline Q$ `commute' with $\overline\cM$ and $\cM$, respectively, in the tensor product, i.e.
\begin{equation}\label{useful id 2}
\begin{split}
\big(Q+\overline{Q}\big)\big(\cM\otimes\overline{\cM}\big)&=\big(Q\,\cM\big)\otimes\overline{\cM}+\cM\otimes\big(\overline{Q}\,\overline{\cM}\big)\;.
\end{split}    
\end{equation}
We start by computing the first term of \eqref{B2 leibniz operator}:
\begin{equation}\label{Leib proof1}
\begin{split}
\Q\,\cB&=-\tfrac12\,\Q\,\cP_\Delta\,b^-\cM\otimes\overline\cM=\tfrac12\,\cP_\Delta\,b^-\big(Q+\overline{Q}\big)\big(\cM\otimes\overline\cM\big)\\
&=\tfrac12\,\cP_\Delta\,b^-\Big\{\big(Q\,\cM\big)\otimes\overline{\cM}+\cM\otimes\big(\overline{Q}\,\overline{\cM}\big)\Big\}\,, \\
\end{split}    
\end{equation}
where we used \eqref{useful id} and \eqref{useful id 2} to move the factors of $Q$ and $\overline Q$ in the tensor product. 
The expression above is meant to act on $\cV\otimes\cV\subset\big(\cK\otimes\overline{\cK}\big)\otimes\big(\cK\otimes\overline{\cK}\big)$, with the unbarred and barred operators in \eqref{Leib proof1} acting on the $\cK\otimes\cK$ and $\overline{\cK}\otimes\overline{\cK}$ subspaces, respectively.
One can now use the Yang-Mills Leibniz relation \eqref{OpLeibniz YM} and write \eqref{Leib proof1} as
\begin{equation}\label{Leib proof2}
\begin{split}
\Q\,\cB&=\tfrac12\,\cP_\Delta\,b^-\Big\{\Big[\cM\big(Q\otimes\mathds{1}+\mathds{1}\otimes Q\big)\Big]\otimes\overline{\cM}+\cM\otimes\Big[\overline{\cM}\big(\overline{Q}\otimes{\mathds{1}}+{\mathds{1}}\otimes \overline{Q}\big)\Big]\Big\}\\
&=\tfrac12\,\cP_\Delta\,b^-\big(\cM\otimes\overline{\cM}\big)\,\Big\{\big(Q\otimes\mathds{1}+\mathds{1}\otimes Q\big)+\big(\overline{Q}\otimes{\mathds{1}}+{\mathds{1}}\otimes \overline{Q}\big)\Big\}\\
&=-\cB\,\Big\{\big(Q+\overline{Q}\big)\otimes\mathds{1}+\mathds{1}\otimes\big({Q}+ \overline{Q}\big)\Big\}\\
&=-\cB\,\big(\Q\otimes\mathds{1}+\mathds{1}\otimes \Q\big)\;,
\end{split}    
\end{equation}
where we used that $\Q\,\cB$ is meant to act on $\cV\otimes\cV$, ensuring that $Q+\overline{Q}=\Q$ in the last line above.

\subsection{Action and Gauge Transformations} 

Having proved that the two-bracket $B_2$ obtained by doubling obeys the Leibniz rule with respect to $\mathds{Q}$, the resulting DFT is guaranteed to be consistent at cubic order and so must be equivalent to the formulation by Hull and Zwiebach \cite{Hull:2009mi}. As a consistency check we explicitly compute the entire cubic action, as well as the gauge transformations of the classical fields, and show that they coincide with the original form in \cite{Hull:2009mi} up to field and parameter 
redefinitions.

We start by identifying the gauge transformations of the fields $\psi\equiv (e_{\mu\bar{\nu}}, e, \bar e, f_\mu, \bar f_{\bar \mu})$. To do so, we compute the gauge bracket by applying the operator \eqref{doubling ansatz} on the tensor product $(\Lambda\otimes\psi)$, using the explicit expressions \eqref{m2 oscillators good}.
We recall that the gauge parameter $\Lambda$ and the string field $\psi$ read
\begin{equation}\label{psi and lambda DFT}
\begin{split}
\Lambda&=\ket{\theta_+\bar\theta_{\bar \mu}}\,\bar\lambda^{\bar \mu}-\ket{\theta_\mu\bar\theta_+}\,\lambda^\mu-2\,\ket{c^+\theta_+\bar\theta_+}\,\eta\;,\\
\psi &=\ket{\theta_\mu\bar\theta_{\bar \nu}}\,e^{\mu\bar{\nu}}+2\,\ket{\theta_+\bar\theta_-}\,\bar e+2\,\ket{\theta_-\bar\theta_+}\, e+2\,\ket{c^+\theta_+\bar\theta_{\bar \mu}}\,\bar f^{\bar \mu}+2\,\ket{c^+\theta_\mu\bar\theta_+}\, f^\mu\;,
\end{split}    
\end{equation}
and the deformed gauge transformations can be read off from $\delta\psi=\mathds{Q}(\Lambda)+\cB(\Lambda\otimes \psi)$.
As an explicit example, let us consider the action of $\cB$ on $\lambda_\mu$ and $e$, i.e. we restrict to $\Lambda=-\ket{\theta_\mu\bar\theta_+}\,\lambda^\mu$ and $\psi=2\,\ket{\theta_-\bar\theta_+}\, e$. One then obtains
\begin{equation}
\begin{split}
\cB\big(\Lambda\otimes\psi\big)&=\cP_\Delta\,b^-\cM\otimes\overline\cM\,\big(\ket{\theta_\mu\bar\theta_+}\,\lambda^\mu\otimes \ket{\theta_-\bar\theta_+}\, e\big)\\
&=\tfrac12\,\cP_\Delta\,\big(b\cM\otimes\overline\cM-\cM\otimes\bar b\,\overline\cM\big)\,\big(\ket{\theta_\mu\bar\theta_+}\,\lambda^\mu\otimes \ket{\theta_-\bar\theta_+}\, e\big)\\
&=\tfrac12\,\cP_\Delta\,\big(b\ket{c\,\theta_-}\,\bra{\theta^{\mu*}}\bra{\theta^{-*}}\del_{2\mu}\big)\otimes\big(\ket{\bar\theta_+}\,\bra{\bar\theta^{+*}}\bra{\bar\theta^{+*}}\big)\big(\ket{\theta_\mu\bar\theta_+}\,\lambda^\mu\otimes \ket{\theta_-\bar\theta_+}\, e\big)\\
&=\tfrac12\,\cP_\Delta\,\ket{\theta_-\bar\theta+}\,\big(\lambda^\mu\del_\mu e\big)\;.
\end{split}    
\end{equation}
One can compute all components of $\cB(\Lambda\otimes \psi)$ in the same fashion, finally yielding
\begin{equation}\label{DFT gauge transf}
\begin{split}
\delta e_{\mu\bar{\nu}} &= \del_\mu\bar\lambda_{\bar \nu}+\bdel_{\bar{\nu}}\lambda_\mu+\tfrac14\,(\lambda\bullet e_{\bar\nu})_{\mu}+\tfrac14\,(\bar\lambda\bullet e_\mu)_{\bar\nu}+\tfrac12\,\lambda_\mu\,\big(\bar f_{\bar\nu}-\bdel_{\bar\nu}\bar e\big)-\tfrac12\,\bar\lambda_{\bar\nu}\,\big(f_\mu-\del_\mu e\big)\;,\\
\delta e &= -\tfrac12\,\del\cdot\lambda+\eta-\tfrac14\,\lambda^\mu\big(f_\mu-\del_\mu e\big)\;,\\
\delta \bar e &=\tfrac12\,\bdel\cdot\bar\lambda+\eta-\tfrac14\,\bar\lambda^{\bar\mu}\big(\bar f_{\bar\mu}-\bdel_{\bar\mu}\bar e\big)\;,\\
\delta f_\mu&=-\tfrac12\,\Box\lambda_\mu+\del_\mu\eta-\tfrac18\,\bdel^{\bar\nu}(\lambda\bullet e_{\bar\nu})_{\mu}+\tfrac14\,\bdel^{\bar\nu}\big[\bar\lambda_{\bar\nu}\, \big(f_\mu-\del_\mu e\big)\big]\;,\\
\delta \bar f_{\bar\mu}&=\tfrac12\,\Box\bar\lambda_{\bar\mu}+\bdel_{\bar\mu}\eta+\tfrac18\,\del^\nu(\bar\lambda\bullet e_\nu)_{\bar\mu}+\tfrac14\,\del^\nu\big[\lambda_\nu\,\big(\bar f_{\bar\mu}-\bdel_{\bar\mu}\bar e\big)\big]\;,
\end{split}    
\end{equation}
where a projection $\cP_\Delta$ is implied on every quadratic term, and we used the bullet product \eqref{bullet}. Specifically, the $\bar\nu$ index in $(\lambda\bullet e_{\bar\nu})_{\mu}$ is viewed as inert, with the $\bullet$ product only acting on unbarred indices. The opposite happens for the $\mu$ index in $(\bar\lambda\bullet e_\mu)_{\bar\nu}$.

One can use the gauge transformations \eqref{DFT gauge transf} to derive the gauge brackets between parameters, upon taking two successive transformations. Since the operator \eqref{doubling ansatz} already contains all the DFT brackets, one can simply compute the gauge algebra by applying \eqref{doubling ansatz} to $\Lambda_1\otimes\Lambda_2$, yielding $B_2(\Lambda_1, \Lambda_2)=\Lambda_{12}$, with
\begin{equation}\label{gauge algebra}
\begin{split}
\lambda_{12}^\mu&=\tfrac14\,(\lambda_1\bullet  \lambda_2)^\mu-\tfrac14\,\bdel_{\bar\nu}\big(\lambda^\mu_1\,\bar\lambda_2^{\bar\nu}\big)+\tfrac14\,\bdel_{\bar\nu}\big(\lambda^\mu_2\,\bar\lambda_1^{\bar\nu}\big)\;,\\
\bar\lambda_{12}^{\bar\mu}&=\tfrac14\,(\bar\lambda_1\bullet  \bar\lambda_2)^{\bar\mu}-\tfrac14\,\del_{\nu}\big(\bar\lambda_1^{\bar\mu}\lambda^\nu_2\big)+\tfrac14\,\del_{\nu}\big(\bar\lambda_2^{\bar\mu}\lambda^\nu_1\big)\;,\\
\eta_{12}&=-\tfrac18\,\del_\mu\bdel_{\bar\nu}\big(\lambda_1^\mu\bar\lambda_2^{\bar\nu}-\lambda_2^\mu\bar\lambda_1^{\bar\nu}\big)\;,
\end{split}    
\end{equation}
where the projector $\cP_\Delta$ is left implicit.
Let us notice that this basis of parameters allows one (to this order) to consistently set $\bar\lambda^{\bar\mu}$ and $\eta$ to zero, in which case \eqref{gauge algebra} reduces to
$B_2^\mu(\lambda_1,  \lambda_2)=(\lambda_1\bullet  \lambda_2)^\mu$. 

The gauge transformations \eqref{DFT gauge transf} do not coincide yet with the ones given in \cite{Hull:2009mi}. In order to match the latter, one has to redefine the gauge parameters as
\begin{equation}
\begin{split}
\lambda_\mu^{\rm HZ}&=\lambda_\mu-\tfrac14\,\bar e\,\lambda_\mu\;,\qquad
\bar\lambda_{\bar\mu}^{\rm HZ}=\bar\lambda_{\bar\mu}+\tfrac14\, e\,\bar\lambda_{\bar\mu}\;,\\
\eta^{\rm HZ}&=\eta+\tfrac{1}{16}\,\eta\,(e-\bar e)-\tfrac18\,(\lambda\cdot\del\bar e+\bar\lambda\cdot\bdel e)-\tfrac{3}{32}\,(\del\cdot\lambda\,\bar e+\bdel\cdot\bar\lambda\,e)\;,
\end{split}    
\end{equation}
and further redefine the scalars as
\begin{equation}\label{HZ field redef e}
e^{\rm HZ}=e-\tfrac{1}{16}\,e\bar e-\tfrac18\,e^2\;,\qquad
\bar e^{\rm HZ}=\bar e+\tfrac{1}{16}\,e\bar e+\tfrac18\,\bar e^2\;.
\end{equation}

Having discussed the gauge transformations, we can now turn to the action. Given a classical string field $\psi$ as in \eqref{psi and lambda DFT} and a dual field equation
\begin{equation}
\cF=\ket{c^+\theta_\mu\bar\theta_{\bar{\nu}}}\,F^{\mu\bar{\nu}}+\ket{c^+\theta_+\bar\theta_-}\,\bar F
+\ket{c^+\theta_-\bar\theta_+}\, F+\ket{\theta_\mu\bar\theta_-}\, F^\mu+\ket{\theta_-\bar\theta_{\bar \mu}}\,\bar F^{\bar \mu}\;,
\end{equation}
the inner product can be defined by the pairing
\begin{equation}
\big\l\psi, \cF\big\r=\int dxd\bar x\,\Big[\tfrac12\,e^{\mu\bar{\nu}}F_{\mu\bar{\nu}}+\bar e F+e\bar F+f^\mu F_\mu+\bar f^{\bar \mu}
\bar F_{\bar \mu}\Big]\;.    
\end{equation}
The action, up to cubic order, is given in $L_\infty$ language by
\begin{equation}
S=\tfrac12\,\big\l\psi, \mathds{Q}(\psi)\big\r +\tfrac{1}{6}\,\big\l\psi, \cB(\psi\otimes \psi)\big\r+\cO(\psi^4)\;,    
\end{equation}
where $\mathds{Q}(\psi)$ and the two-bracket $\cB(\psi\otimes\psi)$ can be computed by using \eqref{Q+Qbar}, \eqref{Q} and \eqref{doubling ansatz}, \eqref{m2 oscillators good}, respectively. After a number of integrations by parts one obtains
\begin{equation}\label{finalCubicCFT}
\begin{split}
S&=\int dxd\bar x\,\Big[\tfrac14\,e^{\mu\bar{\nu}}\square e_{\mu\bar{\nu}}+2\,\bar e\,\square e-f^\mu f_\mu
-\bar f^{\bar \mu} \bar f_{\bar \mu}-f^\mu\Big(\bdel^{\bar \nu} e_{\mu\bar{\nu}}-2\,\del_\mu\bar e\Big)+\bar f^{\bar \nu}
\Big(\del^\mu e_{\mu\bar{\nu}}+2\,\bdel_{\bar \nu} e\Big)\\
&\hspace{+5mm}+\tfrac18\,e^{\mu\bar{\nu}}\Big(\bdel^{\bar \lambda} e_{\mu\bar{\lambda}}\,\del^\rho e_{\rho\bar{\nu}}+\del^\lambda e_{\lambda\bar{\rho}}\,\bdel^{\bar \rho} e_{\mu\bar{\nu}}+2\,\del_\mu e_{\lambda\bar{\rho}}\,\bdel_{\bar \nu} e^{\lambda\bar{\rho}}-2\,\del_\mu e^{\lambda\bar{\rho}}\,\bdel_{\bar \rho} e_{\lambda\bar{\nu}}-2\,\bdel_{\bar{\nu}} 
e^{\lambda\bar{\rho}}\,\del_\lambda e_{\mu\bar{\rho}}\Big)\\
&\hspace{+5mm}+\tfrac12\,e^{\mu\bar{\nu}}\Big(f_\mu-\del_\mu e\Big)\Big(\bar f_{\bar{\nu}}-\bdel_{\bar \nu}\bar e\Big)\Big] \;.   
\end{split}    
\end{equation}
Let us mention that every component of $\cB(\psi\otimes\psi)$ has a projector $\cP_\Delta$, which we assume to be self-adjoint\footnote{In the standard case of a toroidal background, the projector is a Kronecker delta $\cP_\Delta=\delta_{\Delta, 0}$ and the above is clearly obeyed.}, thus yielding the identity when acting on the weakly constrained field $\psi$.
In order to match the action of Hull and Zwiebach \cite{Hull:2009mi}, one has to perform the field redefinition \eqref{HZ field redef e}, together with
\begin{equation}
\begin{split}
f_\mu^{\rm HZ}&=f_\mu-\tfrac14\,\bar e\,f_\mu+\tfrac{3}{16}\,\bar e\,\del_\mu e-\tfrac{1}{16}\,e\,\del_\mu\bar e\;,\\
\bar f_{\bar \mu}^{\rm HZ}&=\bar f_{\bar \mu}+\tfrac14\, e\,\bar f_{\bar \mu}-\tfrac{3}{16}\, e\,\bdel_{\bar \mu} \bar e+\tfrac{1}{16}\,\bar e\,\bdel_{\bar \mu} e \;.   
\end{split}
\end{equation}
We stress that the form (\ref{finalCubicCFT}) of the cubic DFT action is a significant simplification of the one 
given in \cite{Hull:2009mi}.

\section{Conclusions and Outlook}

In this paper we strengthened  the recent results in \cite{Diaz-Jaramillo:2021wtl} according to which 
a natural Lagrangian implementation of double copy applied to Yang-Mills theory yields, at least to cubic order, double field theory. 
More precisely, in \cite{Diaz-Jaramillo:2021wtl} this was shown upon integrating out the DFT dilaton and, for the cubic couplings,  upon choosing Siegel gauge. 
Here we generalized these results by providing a gauge invariant and off-shell   double copy procedure 
that yields the complete DFT to cubic order, including all dilaton couplings. 
To this end we employed a formulation  in terms of strongly homotopy Lie or 
$L_{\infty}$-algebras, which encode the complete data of a classical field theory.
Our results highlight the usefulness of homotopy algebras for a first-principle understanding of double copy. 
(This point was also made at some length in \cite{Borsten:2021hua}.)

We close this section with a brief list of open questions and possible follow-up projects: 
\begin{itemize} 

\item It would be interesting to generalize our construction  to other gauge and DFT-type theories, for instance to 
supersymmetric Yang-Mills theory which should be related to supersymmetric DFT \cite{Hohm:2011nu,Jeon:2011sq}. 

\item Given the relation between the $L_{\infty}$-algebras of Yang-Mills theory and DFT at least to cubic order 
there should be an intimate relation between the classical (perturbative) solutions of both theories. In the homotopy algebra 
formulation these solutions are given by the Maurer-Cartan elements corresponding to the $L_{\infty}$ structure. 
It would be interesting to investigate this  in detail and to compare with existing attempts in the literature 
on establishing double copy relations at the level of classical solutions \cite{Monteiro:2014cda,Luna:2016hge, Monteiro:2021ztt}.

\item 
Arguably the most important  open problem is that of extending this gauge invariant and off-shell 
double copy procedure to quartic and ultimately to all orders. This appears to be  a hard problem for 
the following reason: The 2-bracket of DFT is defined by $\cB=-\tfrac12\,\cP_\Delta\,b^-\cM\otimes\overline\cM$ 
in terms of the 2-product  $\cM$ of the kinematic algebra of Yang-Mills theory, but 
it also involves projectors onto the subspace of 
level-matched states. This projection seems to be  indispensable for various reasons, notably for 
degree reasons, and it leads to significant technical complications when trying to establish the next 
$L_{\infty}$ relation involving the Jacobiator of the 2-bracket and the 3-bracket encoding the quartic 
couplings. Intriguingly, however, solving this problem would also amount to constructing a weakly constrained 
DFT which so far has only been possible to cubic order (due to exactly the same technical 
challenges). Such a weakly constrained DFT is guaranteed to exist, since in principle it is derivable 
from the full closed string field theory by integrating out all string modes except the ones of the DFT sector 
\cite{Sen:2016qap,Arvanitakis:2020rrk,Arvanitakis:2021ecw}. 
Due to the complications of closed string field theory it would, however,  be extremely challenging to do so 
explicitly, and it is an enticing prospect that double copy might provide a short cut.

\end{itemize}

\subsection*{Acknowledgements}

We would like to thank Christoph Chiaffrino, Tomas Codina, Allison Pinto, Jan Plefka and Barton Zwiebach for useful 
discussions and correspondence.

This work is funded   by the European Research Council (ERC) under the European Union's Horizon 2020 research and innovation programme (grant agreement No 771862)
and by the Deutsche Forschungsgemeinschaft (DFG, German Research Foundation), ``Rethinking Quantum Field Theory", Projektnummer 417533893/GRK2575.

\appendix

\section{BRST Quantization of Worldline Theory}\label{worldline}

In this appendix  we show how to derive the formulation of Yang-Mills theory used  in section \ref{sec:LooYM}
by a worldline quantization, using the $\cN=2$ spinning particle \cite{Berezin:1976eg,Gershun:1979fb, Howe:1989vn, Bastianelli:2005vk} and its BRST quantization.

We begin by reviewing the classical worldline theory and then proceed to its BRST quantization.
The fundamental worldline fields constitute a graded phase space which consists of the  bosonic canonical pair $(x^\mu, p_\mu)$, representing target space coordinates and momenta, together with a fermionic canonical pair $(\theta^\mu,\theta^*_\mu)$, which is associated with spin degrees of freedom in spacetime. The fermions $\theta^\mu$ and $\theta_\mu^*$ are related by complex conjugation, i.e.~$(\theta_\mu)^*=\theta^*_\mu$, where we raise and lower spacetime indices with the Minkowski metric.
The rigid model is invariant under global supersymmetries, as well as global time translations, generated by supercharges and the free Hamiltonian, respectively:
\begin{equation}\label{susy charges}
q=\theta^\mu p_\mu\;,\quad q^*=\theta^{\mu*} p_\mu\;,\quad H=\tfrac12\,p^2\;.    
\end{equation}
The $\cN=2$ spinning particle is constructed by gauging worldline supersymmetries and translations, thus turning them into local symmetries. This is achieved by means of complex worldline gravitini $\chi$ and $\chi^*$ and an einbein $e$, respectively, leading to the action
\begin{equation}\label{N2 action}
S=\int d\tau\,\Big[p_\mu\,\del_\tau x^\mu+i\,\theta^*_\mu\,\del_\tau\theta^\mu-\tfrac12\,e\,p^2-i\,\chi^*\,\theta^\mu p_\mu-i\,\chi\,\theta^{\mu*} p_\mu\Big]  \;.  
\end{equation}
The action \eqref{N2 action} is invariant under time reparametrizations $\tau\rightarrow\tau-\xi(\tau)$ and local supersymmetries $\big(\epsilon(\tau), \epsilon^*(\tau)\big)$, with transformations laws
\begin{equation}
\begin{split}
\delta x^\mu &= \xi\,p^\mu+i\,\epsilon\,\theta^{\mu*}+i\,\epsilon^*\theta^\mu\;,\qquad\delta p_\mu=0\;,\\
\delta\theta^\mu&=-\epsilon\,p^\mu\;,\qquad \hspace{24mm}\delta\theta^{\mu*}=-\epsilon^*p^\mu\;,\\
\delta e&=\del_\tau\xi+2i\,\epsilon\,\chi^*+2i\,\epsilon^*\chi\;,\qquad\hspace{1mm}\delta\chi=\del_\tau\epsilon\;,\qquad\delta\chi^*=\del_\tau{\epsilon^*}\;,
\end{split}
\end{equation}
where both $\epsilon^*$ and $\chi^*$ are related to $\epsilon$ and $\chi$ by complex conjugation.
Canonical quantization of the graded symplectic structure of \eqref{N2 action} gives rise to the (anti)-commutation relations
\begin{equation}\label{canconmu}
[x^\mu, p_\nu]=i\,\delta^\mu{}_\nu\;,\quad \{\theta^\mu, \theta^*_\nu\}=\delta^\mu{}_\nu\;,    
\end{equation}
where we denoted the quantum operators with the same symbols as the classical variables, since from now on we will only work with the quantum theory.
The generators of local (super)symmetries \eqref{susy charges} turn into quantum first-class constraints obeying the superalgebra
\begin{equation}\label{susy algebra}
\begin{split}
\{q, q\}&=0\;,\quad  \{q^*, q^*\}=0\;,\quad\{q,  q^*\}=2\,H\;,\\    
[H,q]&=0\;,\quad\hspace{2mm}[H, q^*]=0\;.
\end{split}    
\end{equation}

The worldline model, as it stands, describes abelian massless $p-$forms of arbitrary degree. In order to accommodate color degrees of freedom, we shall extend the graded phase space by a further fermionic canonical pair $(t_a, t^{a*})$, as originally introduced in \cite{Bastianelli:2013pta}, and supplement the action \eqref{N2 action} by the term
\begin{equation}
S_{\rm color}=i\int d\tau \,t^{a*}\,\del_\tau t_a \;,   
\end{equation}
where $a=1,\cdots,{\rm dim}\,\mathfrak{g}$ is an adjoint index of a Lie algebra $\mathfrak{g}$. Canonical quantization of these fermions leads to the anticommutator
\begin{equation}
\{t_a, t^{b*}\}=\delta_a{}^b\;,    
\end{equation}
and we shall lower and raise adjoint indices with the Cartan-Killing metric $\kappa_{ab}=\delta_{ab}$ and its inverse.
Rather than studying the physical state conditions of the theory in this form, we shall proceed to its BRST quantization, which makes the spacetime gauge structure manifest.

\paragraph{BRST-extended Hilbert space}

The canonical quantization of the action \eqref{N2 action} leads to a constrained Hamiltonian system, with first-class quantum constraints $(q, q^*, H)$. In order to proceed with the BRST formalism we introduce a canonical (super)ghost pair for each constraint, namely
\begin{equation}
H\;\rightarrow\;(b,c)\;,\quad q\;\rightarrow\;(\beta,\gamma^*)\;,\quad q^*\;\rightarrow\;(\beta^*,\gamma)\;,    
\end{equation}
with Grassmann parity $\epsilon$ and ghost number assignments as follows:
\begin{equation}\label{wl degrees}
\epsilon(b,c)=1\;,\quad\epsilon(\gamma, \gamma^*, \beta, \beta^*)=0\;,\quad{\rm gh}(c, \gamma, \gamma^*)=+1\;,\quad{\rm gh}(b, \beta, \beta^*)=-1  \;,  
\end{equation}
while all other worldline operators have ghost number zero.
The canonical (anti)commutation relations are given by
\begin{equation}\label{cancon+-}
\{b,c\}=1\;,\quad[\beta,\gamma^*]=1\;,\quad[\beta^*, \gamma]=1\;,
\end{equation}
with all other (anti)commutators vanishing. Coming now to the extended Hilbert space, we realize the $(x,p)$ algebra as usual by identifying $p_\mu=-i\,\del_\mu$ acting on smooth functions of $x$. The space of smooth functions of $x$ is then tensored with the Fock space of matter (the $\theta^\mu$ and $t_a$ sectors) and ghost oscillators. We choose the Fock vacuum to be annihilated by $b$ and all starred operators:
\begin{equation}
\big(b,\theta^*_\mu, t^{a*},\gamma^*, \beta^*\big)\lvert0\rangle=0\;.  \end{equation}
A basis of the Fock space is thus given by arbitrary monomials in all the creation operators acting on the vacuum $\ket{0}$, which we assume to be bosonic and of ghost number zero:
\begin{equation}\label{Fock basis}
\ket{i,j,k|^{\mu_1\cdots\mu_p}_{a_1\cdots a_q}}=(\gamma)^i(\beta)^j(c)^k(\theta^{\mu_1}\cdots\theta^{\mu_p})\,(t_{a_1}\cdots t_{a_q})\,\ket{0}  \;.  
\end{equation}
An arbitrary state of the Hilbert space is thus given by linear combinations of smooth functions tensored with the states \eqref{Fock basis}:
\begin{equation}\label{state}
\ket{\psi}=\sum_{p=0}^D\sum_{q=0}^{{\rm dim}\,\mathfrak{g}}\sum_{i,j=0}^\infty\sum_{k=0}^1\ket{i,j,k|^{\mu_1\cdots\mu_p}_{a_1\cdots a_q}}\, \psi_{\mu_1\cdots\mu_p}^{(ijk)\,a_1\cdots a_q}(x)\;,   \end{equation}
where the functions of $x$ are interpreted as spacetime $p-$forms taking values in antisymmetric products of the adjoint representation of $\mathfrak{g}$. Using the definition \eqref{Fock basis} of the Fock space basis, one can sum over the oscillator numbers in \eqref{state} and rewrite the arbitrary state $\ket{\psi}$ as
\begin{equation}\label{state smart}
\ket{\psi}=\Psi(x,\theta, t|\gamma, \beta;c)\ket{0}  \;,  
\end{equation}
where the $\Psi$ on the right-hand side is taken to be an operator-valued function acting on the vacuum.
This correspondence is analogous to the operator-state correspondence in string theory and
allows one to identify the state $\ket{\psi}$ itself with the operator $\Psi$. In the following, we will refer interchangeably to the state $\ket{\psi}$ and the operator $\Psi$ as the `string field'.

\paragraph{Truncation and Yang-Mills complex}

Turning  to the spectrum of off-shell states, one infers  from \eqref{state} that it goes vastly beyond the spectrum of Yang-Mills. In order to remedy this, we shall decompose the full Hilbert space $\cH$ by means of two number operators:
\begin{equation}\label{YM numbers}
\begin{split}
\cN&=\theta^\mu\theta^*_\mu+\gamma\beta^*-\beta\gamma^*=N_\theta+N_\gamma+N_\beta\;,\\
N_t&=t_a\,t^{a*}\;,
\end{split}    
\end{equation}
whose choice will be clarified in the following.
The Hilbert space thus decomposes as a double direct sum as follows:
\begin{equation}\label{H decomp}
\cH=\bigoplus_{\cN=0}^\infty\bigoplus_{N_t=0}^{{\rm dim}\,\mathfrak{g}}\cH_{\cN,N_t} \;,   
\end{equation}
where in the above formula we denoted the eigenvalues with the same symbol as the operators \eqref{YM numbers}. The basis elements \eqref{Fock basis}, for instance, belong to the subspaces $\cH_{m,n}$ according to $\ket{i,j,k|^{\mu_1\cdots\mu_p}_{a_1\cdots a_q}}\in\cH_{p+i+j, q}$. The Hilbert subspace describing Yang-Mills is given by $\cH_{1,1}$ which, as we will now show, is isomorphic to the $L_\infty$ complex $\cX$.

Let us study in more detail the structure of the subspace $\cH_{1,1}$. To this end, and to make contact with the formulation of section \ref{sec:SFT YM}, it is convenient to group the creation operators $(\theta^\mu, \gamma, \beta)$ and the annihilation operators $(\theta^{\mu*}, \gamma^*, \beta^*)$ into the graded oscillators
\begin{equation}\label{wl thetas}
\theta_M=(\theta_+, \theta_\mu, \theta_-)=(-i\,\beta, \theta_\mu, -i\,\gamma)\;,\quad\theta^{M*}=(\theta^{+*}, \theta^{\mu*}, \theta^{-*})=(-i\,\gamma^*, \theta^{\mu*}, i\,\beta^*)\;.    
\end{equation}
The ghost number assignments \eqref{wl degrees} and graded commutation relations \eqref{canconmu}, \eqref{cancon+-} can be summarized as
\begin{equation}
{\rm gh}\big(\theta_M\big)=-M\;,\quad{\rm gh}\big(\theta^{M*}\big)=M\;,\quad[\theta^{M*}, \theta_N\}=\delta^M{}_N\;,
\end{equation}
which will be useful in identifying the Yang-Mills $L_\infty$ complex as $\cX\simeq\cH_{1,1}$. As in section \ref{sec:SFT YM}, the index $M=(+,\mu, -)$ counts as $(+1, 0, -1)$ when assigning degrees.
By using the creation operators $\theta_M$ and $c$ one can introduce the basis for the Fock space of $\cH_{1,1}$ as
\begin{equation}
\ket{\theta_Mt_a}=\theta_Mt_a\ket{0}\;,\quad \ket{c\,\theta_Mt_a}=c\,\theta_Mt_a\ket{0}\;.   
\end{equation}
One can now write an arbitrary state of $\cH_{1,1}$ in a compact way as
\begin{equation}\label{H11 vector}
\ket{u}=\ket{\theta_M t_a}\,u^{Ma}(x)+\ket{c\,\theta_M t_a}\,v^{Ma}(x)\,\in\cH_{1,1}\;,    
\end{equation}
which, comparing with \eqref{general decomp K} and \eqref{general decomp X}, makes it clear that $\cH_{1,1}$ is isomorphic to the $L_\infty$ complex of Yang-Mills $\cX$. The $L_\infty$ degree assignments \eqref{general decomp X} are recovered, once we identify the degree in $\cX$ with (minus) the worldline ghost number:
\begin{equation}
|u|_\cX=-{\rm gh}\big(\ket{u}\big)\;,\quad u\in\cX\;,\quad\ket{u}\in\cH_{1,1}\;,\quad \cX\simeq\cH_{1,1}\;. 
\end{equation}
Similarly, it is clear from \eqref{thetas}, \eqref{cthetas} and the decomposition \eqref{H decomp} that the $C_\infty$ algebra $\cK$ is isomorphic to the Hilbert subspace $\cH_{1,0}$, since the decomposition in kinematic and color degrees of freedom is manifest on the worldline. An arbitrary state of $\cH_{1,0}$ is given, in fact, by
\begin{equation}
\ket{u}=\ket{\theta_M}\,u^{M}(x)+\ket{c\,\theta_M}\,v^{M}(x)\,\in\cH_{1,0}\;,    
\end{equation}
as in \eqref{general decomp K}, with the obvious definition for the vectors $\ket{\theta_M}$ and $\ket{c\,\theta_M}$ 
of the Fock basis of $\cH_{1,0}$. The $C_\infty$ degree is similarly related to the worldline ghost number as
\begin{equation}
|u|_\cK=-{\rm gh}\big(\ket{u}\big)-1\;,\quad u\in\cK\;,\quad\ket{u}\in\cH_{1,0}\;,\quad \cK\simeq\cH_{1,0}\;. 
\end{equation}
Both subspaces $\cH_{1,1}$ and $\cH_{1,0}$ can be further decomposed according to the ghost number, with the aforementioned isomorphisms extending to the subspaces of fixed degree (or ghost number):
\begin{equation}
\begin{split}
\cH_{1,1}&=\bigoplus_{k=-1}^2\big(\cH_{1,1}\big)_k\;,\quad{\rm gh}\big(\cH_{1,1}\big)_k=k\;,\quad \big(\cH_{1,1}\big)_k\simeq X_{-k}\;,\\
\cH_{1,0}&=\bigoplus_{k=-1}^2\big(\cH_{1,0}\big)_k\;,\quad{\rm gh}\big(\cH_{1,0}\big)_k=k\;,\quad \big(\cH_{1,0}\big)_k\simeq K_{-1-k} \;.   
\end{split}    
\end{equation}
Finally, the $\mathbb{Z}_2$ split introduced in section \ref{sec:SFT YM} is nothing but the two-dimensional Hilbert space representing the $(b,c)$ algebra, with the $c-$degree counted by the number operator $N_c=c\,b$.

\paragraph{BRST differential}

After identifying the vector spaces $\cX\simeq\cH_{1,1}$ and $\cK\simeq\cH_{1,0}$, we now turn to the construction of the differential. Given the constraint superalgebra \eqref{susy algebra} and the ghost commutation relations \eqref{cancon+-}, the nilpotent BRST operator can be defined in the standard way, yielding
\begin{equation}
Q=-2\,c\,H+\gamma\,q^*+\gamma^*q+\gamma\,\gamma^*b\;,\quad Q^2=0\;,\quad {\rm gh}\big(Q\big)=+1\;.  
\end{equation}
At this point, we can justify our choice of decomposition \eqref{H decomp} according to the number operators defined in \eqref{YM numbers}: the BRST operator $Q$ commutes with both $\cN$ and $N_t$, making it a well-defined endomorphism on each subspace $\cH_{m,n}$ \cite{Dai:2008bh, Bastianelli:2019xhi, Bonezzi:2020jjq}:
\begin{equation}
\cH=\bigoplus_{m=0}^{\infty}\bigoplus_{n=0}^{{\rm dim}\mathfrak{g}}\cH_{m,n}\;,\quad Q\;:\;\cH_{m,n}\;\longrightarrow\;\cH_{m,n}\;,
\end{equation}
which allows us to study the BRST cohomology on each $\cH_{m,n}$ and, in particular, on the spaces of interest $\cH_{1,1}$ and $\cH_{1,0}$. Since the BRST cohomology only probes the free theory, which is independent of color, we shall focus on the $C_\infty$ complex $\cK\simeq \cH_{1,0}$.
Upon identifying the momentum operator as $p_\mu=-i\,\del_\mu$ and using the definition \eqref{wl thetas}, $Q$ can be rewritten as
\begin{equation}\label{wl Q}
Q=c\,\B+\Big(\theta^\mu\theta^{+*}+\theta_-\theta^{\mu*}\Big)\del_\mu-\theta_-\theta^{+*}b\;,    
\end{equation}
which reproduces the form \eqref{Q} given in section \ref{sec:SFT YM}. In order to see that \eqref{wl Q} actually coincides with the differential \eqref{Q}, it is useful to introduce the bra states for the Fock space of the dual $\cH^*_{1,0}$. To do this, we introduce the bra vacuum state $\bra{0}$, which is annihilated by the $\theta_M$ and $b$:
$\bra{0}\theta_M=\bra{0}b=0$. The fact that the bra vacuum is also annihilated by $b$ is compatible with $(b,c)^\dagger=(b,c)$. This, together with the anticommutator \eqref{cancon+-}, implies the basic overlaps
\begin{equation}\label{overlap}
\bra{0}c\ket{0}=1   \;,\quad\l0|0\r=0\;.
\end{equation}
Given the basis vectors of the Fock space of $\cH_{1,0}$:
\begin{equation}
\ket{\theta_M}=\theta_M\ket{0}\;,\quad \ket{c\,\theta_M}=c\,\theta_M\ket{0}\;,    
\end{equation}
one can construct the dual vectors $\bra{\theta^{M*}}$ and $\bra{\theta^{M*}b}$ introduced in \eqref{bra ket} by defining
\begin{equation}
\bra{\theta^{M*}}=\bra{0}c\,\theta^{M*}\;,\quad \bra{\theta^{M*}b}=\bra{0}c\,\theta^{M*}b \;,
\end{equation}
which allows us to decompose the identity on the Fock space of $\cH_{1,0}$ as
\begin{equation}
\mathds{1}=\ket{\theta_M}\bra{\theta^{M*}}+\ket{c\,\theta_M}\bra{\theta^{M*}b}\;.    
\end{equation}
Inserting the decomposition of the identity in \eqref{wl Q} one recovers the differential in the form \eqref{Q}. For instance, one has
\begin{equation}
\begin{split}
\theta_\mu\theta^{+*}\del^\mu&=\theta_\mu\theta^{+*}\Big(\ket{\theta_M}\bra{\theta^{M*}}+\ket{c\,\theta_M}\bra{\theta^{M*}b}\Big)\del^\mu\\
&=\Big(\ket{\theta_\mu}\bra{\theta^{+*}}-\ket{c\,\theta_\mu}\bra{\theta^{+*}b}\Big)\del^\mu\;,
\end{split}    
\end{equation}
which is the second term of \eqref{Q}. This ensures that the action of $Q$ on $\cH_{1,0}$ reproduces the differential $m_1$ \eqref{allm1s} on $\cK$,
with the different components \eqref{field decomposition} of $\cK$ identified as
\begin{equation}
\ket{\Lambda}\in\big(\cH_{1,0}\big)_{-1}\;,\quad \ket{\cA}\in\big(\cH_{1,0}\big)_{0}\;,\quad \ket{\cE}\in\big(\cH_{1,0}\big)_{1}\;,\quad \ket{\cN}\in\big(\cH_{1,0}\big)_{2}\;.    
\end{equation}

As we have discussed in section \ref{sec:LooYM}, the field theoretic interpretation of field equations, gauge transformations and so on is properly encoded in the $L_\infty$ algebra $\cX$, rather than the kinematic algebra $\cK$. In this respect, given the identification $\cX\simeq\cH_{1,1}$, on-shell fields obeying $b_1(\cA)=0$, modulo gauge symmetries $\delta\cA=b_1(\Lambda)$
are given by the BRST cohomology on $\cH_{1,1}$ at ghost number zero:
\begin{equation}
\begin{split}
Q\ket{\cA}&=0\;,\quad\hspace{6mm}\ket{\cA}\in\big(\cH_{1,1}\big)_0\;\hspace{2mm}\longrightarrow\;\left\{\begin{array}{r}
\B A_\mu{}^a-\del_\mu\varphi^a=0  \\
\del\cdot A^a-\varphi^a=0 
\end{array}\right.\;,\\
 \delta\ket{\cA}&=Q\ket{\Lambda}\;,\quad\ket{\Lambda}\in\big(\cH_{1,1}\big)_{-1}\;\longrightarrow\;\left\{\begin{array}{l}
     \delta A_\mu{}^a=\del_\mu\lambda^a  \\
     \hspace{2mm}\delta\varphi^a=\B\lambda^a
\end{array}\right.    \;.   
\end{split}
\end{equation}


\providecommand{\href}[2]{#2}\begingroup\raggedright\endgroup

\end{document}